%% file: bare_jrnl_new_sample4.tex
\documentclass[journal]{IEEEtran}
\usepackage{amsmath,amsfonts}
\usepackage{array}
\usepackage[caption=false,font=normalsize,labelfont=sf,textfont=sf]{subfig}
\usepackage{textcomp}
\usepackage{stfloats}
\usepackage{url}
\usepackage{graphicx}
\usepackage{cite}
\usepackage[ruled,vlined,linesnumbered]{algorithm2e}

\usepackage[utf8]{inputenc} % allow utf-8 input
\usepackage[T1]{fontenc}    % use 8-bit T1 fonts
\usepackage{hyperref}       % hyperlinks
\usepackage{url}            % simple URL typesetting
\usepackage{booktabs}       % professional-quality tables
\usepackage{amsfonts}       % blackboard math symbols
\usepackage{nicefrac}       % compact symbols for 1/2, etc.
\usepackage{microtype}      % microtypography

\usepackage{times}
\usepackage{epsfig}
\usepackage{graphicx}
\usepackage{amsmath}
\usepackage{amssymb}
\usepackage{multirow}
\usepackage{bbding}
\usepackage{pifont}
\usepackage{makecell}
\usepackage{capt-of}
\usepackage{tabularx}
\usepackage[table]{xcolor}
\usepackage{wrapfig}
\usepackage{enumitem}
\usepackage{bm}

\newcommand{\ie}{\emph{i.e.},~}

\usepackage{fontawesome5}
\usepackage{xcolor}

\newcommand{\frozenicon}{\textcolor{cyan}{\faSnowflake}}
\newcommand{\trainableicon}{\textcolor{orange}{\faFire}}

\usepackage{color}
\definecolor{citecolor}{RGB}{66,168,235}
\definecolor{linkcolor}{RGB}{255,0,0}

\hypersetup{colorlinks=true,citecolor=citecolor,linkcolor=linkcolor}

\definecolor{detcolor}{gray}{.9}

\definecolor{bestcolor}{gray}{.9}

\newlength\savewidth

\begin{document}

\title{Physically Plausible Human-Object Rendering from Sparse Views \\ via 3D Gaussian Splatting}

\author{
\thanks{*Long Chen is the corresponding author.}

Weiquan Wang, Jun Xiao, Yi Yang,~\IEEEmembership{Fellow,~IEEE,} Yueting Zhuang,~\IEEEmembership{Senior Member,~IEEE,} \\ and Long Chen*,~\IEEEmembership{Member,~IEEE} 
        % <-this % stops a space
\thanks{Weiquan Wang, Jun Xiao, Yi Yang, and Yueting Zhuang are with the College of
Computer Science, Zhejiang University, Hangzhou 310027, China (e-mail:
wqwangcs@zju.edu.cn; junx@cs.zju.edu.cn; yangyics@zju.edu.cn; yzhuang@zju.edu.cn).}% <-this % stops a space
\thanks{Long Chen is with the Department of Computer Science and Engineering,
The Hong Kong University of Science and Technology, Clear Water Bay, Hong Kong (e-mail: longchen@ust.hk).}
}

\markboth{Journal of \LaTeX\ Class Files,~Vol.~14, No.~8, August~2021}%
{Shell \MakeLowercase{\textit{et al.}}: A Sample Article Using IEEEtran.cls for IEEE Journals}

\IEEEpubid{0000--0000/00\$00.00~\copyright~2021 IEEE}

\maketitle

\begin{abstract}
Rendering realistic human-object interactions (HOIs) from sparse-view inputs is a challenging yet crucial task for various real-world applications. Existing methods often struggle to simultaneously achieve high rendering quality, physical plausibility, and computational efficiency. To address these limitations, we propose \textbf{HOGS} (\textbf{H}uman-\textbf{O}bject Rendering via 3D \textbf{G}aussian \textbf{S}platting), a novel framework for efficient HOI rendering with physically plausible geometric constraints from sparse views. 
HOGS represents both humans and objects as dynamic 3D Gaussians. Central to HOGS is a novel optimization process that operates directly on these Gaussians to enforce geometric consistency (i.e., preventing inter-penetration or floating contacts) to achieve physical plausibility. 
To support this core optimization under sparse-view ambiguity, our framework incorporates two pre-trained modules: an optimization-guided \textbf{Human Pose Refiner} for robust estimation under sparse-view occlusions, and a \textbf{Human-Object Contact Predictor} that efficiently identifies interaction regions to guide our novel contact and separation losses.
Extensive experiments on both human-object and hand-object interaction datasets demonstrate that HOGS achieves state-of-the-art rendering quality and maintains high computational efficiency.
\end{abstract}

\begin{IEEEkeywords}
Human-object interactions, 3D Gaussian Splatting, sparse-view rendering, \textbf{physically plausible optimization}.
\end{IEEEkeywords}

\input{sec/1_intro}
\input{sec/2_related}

\input{sec/3_method}
\input{sec/4_exp}
\input{sec/5_conclusion}

\section*{Acknowledgments}

This work was supported by the Fundamental and Interdisciplinary Disciplines Breakthrough Plan of the Ministry of Education of China (JYB2025XDXM103), the Key R\&D Program of Zhejiang (2025C01128), the National Natural Science Foundation of China Young Scholar Fund Category B (62522216), Young Scholar Fund Category C (62402408), the National Natural Science Foundation of China (62441617), Zhejiang Provincial Natural Science Foundation of China (No. LD25F020001), the Hong Kong SAR RGC General Research Fund (16219025), and Early Career Scheme (26208924).

\bibliographystyle{IEEEtran}
\bibliography{main}

\begin{IEEEbiography}[{\includegraphics[width=1in,height=1.25in,clip,keepaspectratio]{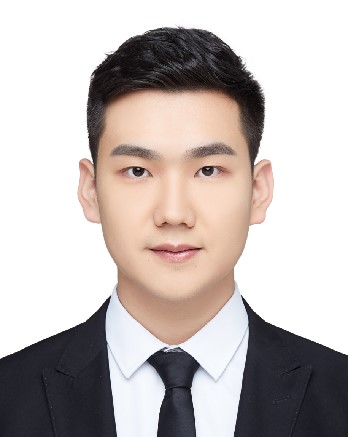}}]{Weiquan Wang}
received the B.S. and M.S. degrees in information and communication engineering from the Harbin Institute of Technology, Harbin, China, in 2021 and 2023, respectively.  
He is currently working toward the PhD degree with the College of Computer Science at Zhejiang University, Hangzhou, China. His current research interests include deep learning and computer vision.
\end{IEEEbiography}

\begin{IEEEbiography}[{\includegraphics[width=1in,height=1.25in,clip,keepaspectratio]{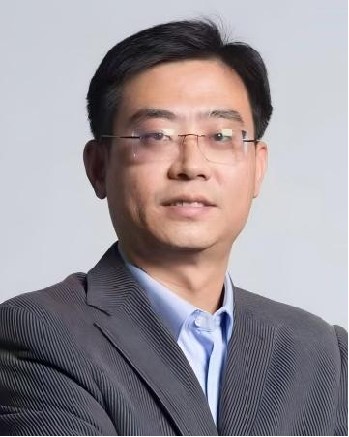}}]{Jun Xiao}
received the Ph.D.degree in computer science and technology from the College of Computer Science, Zhejiang University, Hangzhou, China, in 2007. He is currently a Professor with the College of Computer Science, Zhejiang University. His current research interests include computer animation, multimedia retrieval, and machine learning.
\end{IEEEbiography}

\begin{IEEEbiography}[{\includegraphics[width=1in,height=1.25in,clip,keepaspectratio]{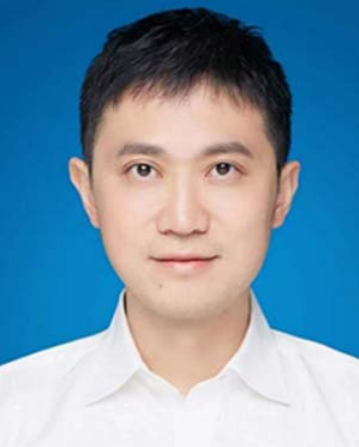}}]{Yi Yang}
(Fellow, IEEE) received the Ph.D degree from Zhejiang University, China, in 2010. He is currently a Distinguished Professor with Zhejiang University. He was a post-doctoral researcher with the School of Computer Science, Carnegie Mellon University. His current research interests include machine learning and multimedia content analysis, such as multimedia retrieval and video content understanding. He received the Australia Research Council Early Career Researcher Award, Computing Society, the Google Faculty Australia Research Award, and the AWS Machine Learning Research Award Gold Disruptor Award.
\end{IEEEbiography}

\begin{IEEEbiography}[{\includegraphics[width=1in,height=1.25in,clip,keepaspectratio]{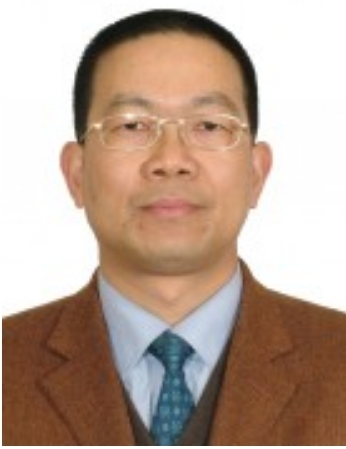}}]{Yueting Zhuang}
(Senior Member, IEEE) received the B.Sc., MSc., and Ph.D. degrees in computer science from Zhejiang University, China, in 1986, 1989, and 1998, respectively. From February 1997 to August 1998, he was a Visiting Scholar with the University of Illinois at Urbana Champaign. He was the Dean of the College of Computer Science, Zhejiang University, from 2009 to 2017; and the Director of the Institute of Artificial Intelligence from 2006 to 2015. He was a CAAl Fellow in 2018 and serves as a Standing Committee Member for CAAl. He was a fellow of the China Society of Image and Graphics in 2019. Also, he is a member of the Zhejiang Provincial Government AI Development Committee (AI Top 30).
\end{IEEEbiography}

\begin{IEEEbiography}[{\includegraphics[width=1in,height=1.25in,clip,keepaspectratio]{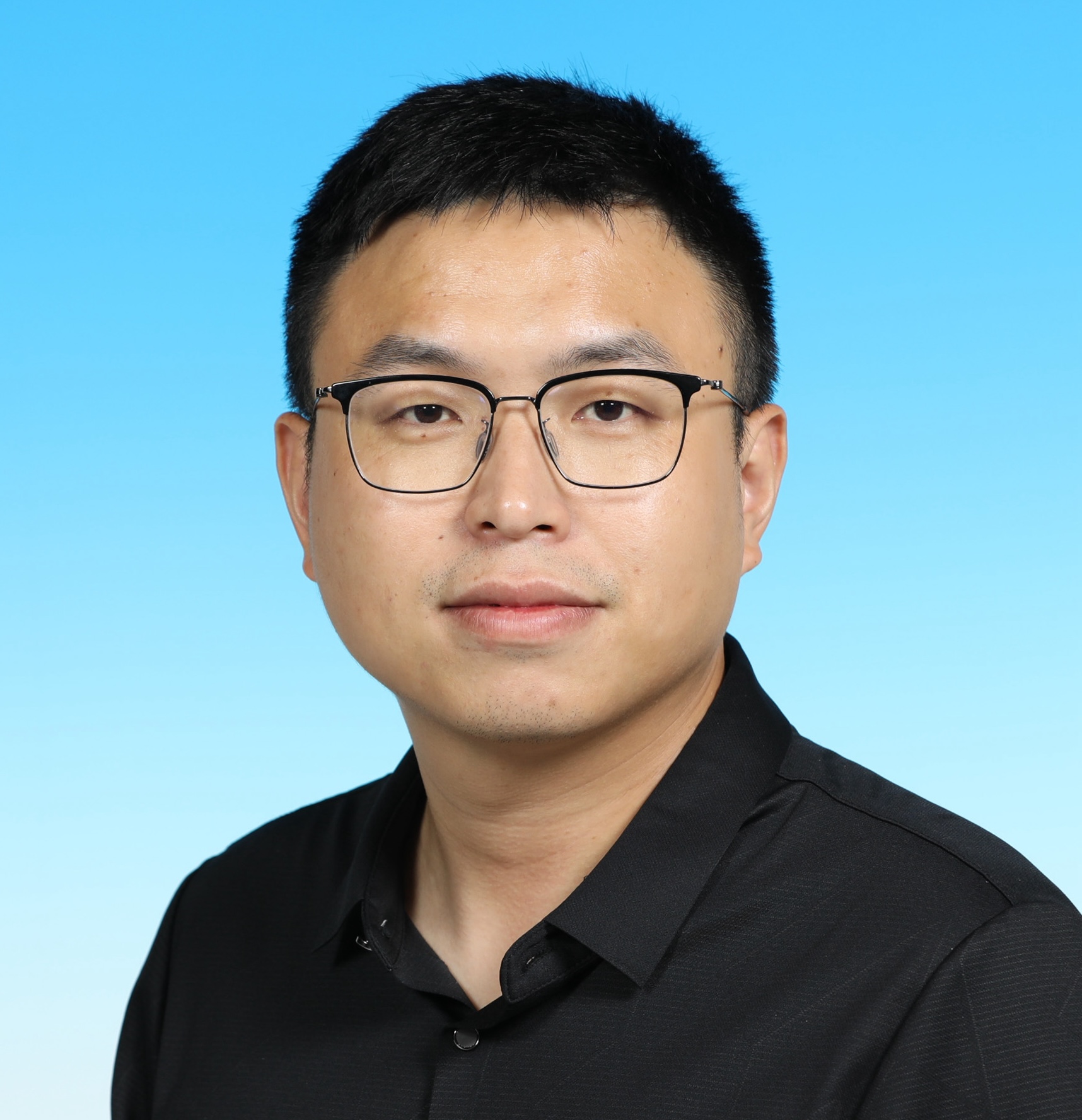}}]{Long Chen}
(Member, IEEE) received the BEng degree in electrical information engineering from the Dalian University of Technology, in 2015, and the PhD degree in computer science from Zhejiang University, in 2020. He is currently an Assistant Professor with The Hong Kong University of Science and Technology (HKUST). He was a postdoctoral research scientist with Columbia University and a senior researcher with Tencent AI Lab. His research interests are computer vision, machine learning, and multimedia.
\end{IEEEbiography}

% \vfill

\end{document}

%% file: sec/1_intro.tex
\section{Introduction}
\label{sec:intro}

Rendering dynamic real-world scenes is crucial for applications ranging from immersive media to robotic interaction~\cite{huang2022capturing, liu2025revisit, li2024task, kong2022human, liu2024citygaussian, kong2025efficient}. 
A fundamental challenge in deploying these applications, however, is the reliance on sparse-view inputs~\cite{zhao2024m, pang2024sparse, li2023multi} (6 or fewer fixed camera views covering a 360° circle), as dense multi-view camera setups are often impractical in real-world environments.
This limitation becomes particularly pronounced when rendering human-object interactions (HOIs), where the standard for realism extends beyond simple visual fidelity~\cite{fernandez2020associated, tonderski2024neurad, boos2016flashback, qiao2019web}. 
For these specific scenes, achieving physical plausibility is paramount. 
Consider the need to verify that a user's avatar is sitting correctly on a virtual chair rather than sinking into it for an immersive AR experience, or to confirm a driver's hand is making proper contact with the steering wheel for a safety analysis. 
Such scenarios demand a level of physical realism that general-purpose renderers struggle to provide, especially given the inherent ambiguity of sparse-view inputs. 
Therefore, jointly achieving high-quality rendering and robust physical plausibility for HOIs from sparse views remains a pivotal and challenging goal.

Meanwhile, HOI rendering has witnessed significant progress within recent years~\cite{jiang2022neuralhofusion, jiang2023instant, bhatnagar2022behave, suo2021neuralhumanfvv}. 
Early approaches rely on 3D mesh reconstruction combined with per-frame texture mapping~\cite{schonberger2016structure, collet2015high, dou2017motion2fusion}, often incorporating physical optimizations to enforce plausible contact between the reconstructed human and object meshes (\ie \textbf{\emph{physically plausible}}). 
While they provide basic visual fidelity, they are susceptible to occlusions and incomplete textures. 
\IEEEpubidadjcol
These challenges are further exacerbated in sparse-view settings, which limit their effectiveness in real scenarios. 
In light of these issues, more advanced rendering techniques are proposed to enhance HOI rendering, among which neural rendering techniques~\cite{su2021nerf, jiang2023instant, sun2021neural, liu2023hosnerf} demonstrate significant gains in \textbf{\emph{rendering quality}}.
A common strategy for neural HOI rendering uses a layer-wise NeRF pipeline to represent and render both human and object~\cite{jiang2022neuralhofusion, zhang2023neuraldome}, enabling free-viewpoint rendering. 
Despite the impressive visual fidelity achievable, this type of approach inherently demands dense multi-view inputs and incurs substantial computational overhead, limiting its practicality for real-time applications.  
Moreover, even with the adoption of layered representations to disentangle human and object, these pipelines generally lack dedicated mechanisms to ensure physically plausible interactions in intricate HOI scenarios.

\begin{figure*}[t]
\begin{center}
% \fbox{\rule{0pt}{2in} \rule{0.98\linewidth}{0pt}}
\includegraphics[width=0.95\linewidth]{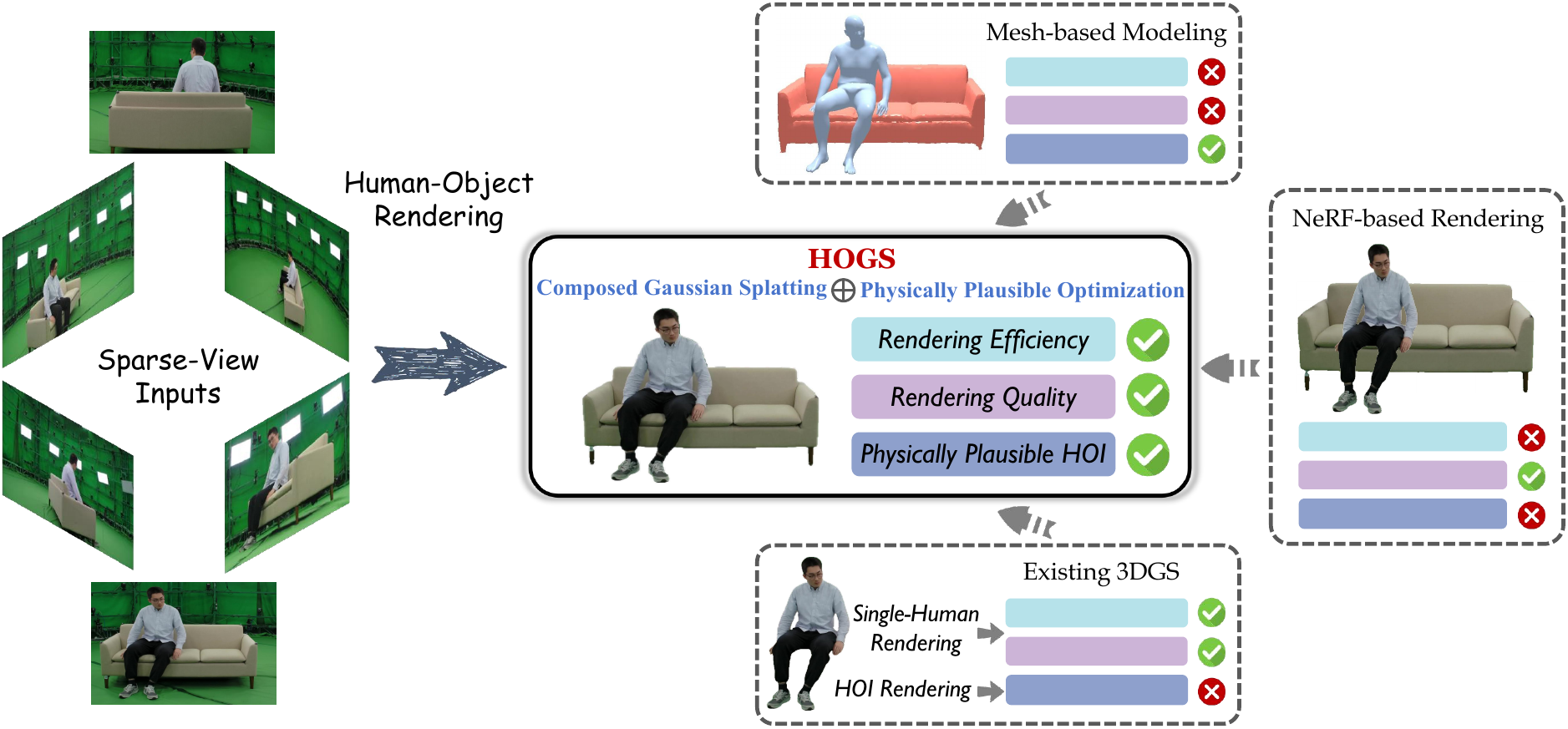}
\end{center}
\caption{\textbf{Comparison of state-of-the-art sparse-view HOI rendering methods.} Mesh-based methods have limitations in rendering efficiency and visual quality. NeRF-based methods, while capable of high fidelity, typically struggle with rendering efficiency and physically plausible HOI. Moreover, existing 3DGS methods lack effective HOI handling despite decent single-human rendering performance. In contrast, our proposed HOGS significantly improves high-quality, efficiency, and physically plausible HOI rendering simultaneously.}
\label{fig:teaser}
% \vspace{-1.5em}
\end{figure*}

To tackle the computational limitations of NeRF-based methods, recent research has explored 3D Gaussian Splatting (3DGS)~\cite{kerbl20233d} as a powerful alternative, offering remarkable \textbf{\emph{rendering efficiency}} for photo-realistic rendering of both static and dynamic scenes~\cite{feng2025flashgs, fei20243d, shaw2025swings, 11024125}.
This explicit Gaussian representation provides robustness to sparse views by efficiently integrating limited information and minimizing reconstruction ambiguity~\cite{zhang2025cor, paliwal2025coherentgs, chen2025mvsplat, mihajlovic2025splatfields}. 
Building upon these strengths, 3DGS has been applied to dynamic human modeling, achieving high-quality reconstruction of animatable human avatars~\cite{liu2024humangaussian, kocabas2024hugs, hu2024gauhuman, moon2024expressive, qiu2025anigs}.
Despite these advancements, existing 3DGS-based methods predominantly focus on single-human rendering. Applying them directly to complex interactions between humans and objects reveals specific limitations due to severe occlusions and geometric ambiguity.

In this paper, we propose \textbf{HOGS} (\textbf{H}uman-\textbf{O}bject Rendering via 3D \textbf{G}aussian \textbf{S}platting), a novel framework explicitly designed to address the specific challenges in sparse-view HOI rendering.
The rationale behind our framework design follows a logical progression to address the limitations of standard 3DGS in this domain:

First, in our sparse-view setting, geometric information is severely missing. Without strong priors, standard 3DGS tends to produce "floaters" or fails to converge accurately. To address this, we introduce a \textbf{Human-Object Deformation} module that leverages parametric priors (SMPL-H~\cite{romero2022embodied} and templates) to provide a robust geometric initialization. This step is a prerequisite for 3DGS to work under sparse views, ensuring the optimization starts from a plausible geometric foundation.

Second, following deformation, we obtain initialized points from two fundamentally different sources: the SMPL-based human (articulated) and the template-based object (rigid). Treating them as a single homogeneous Gaussian cloud is problematic because they require distinct densification and pruning strategies due to their differing dynamic properties. Conversely, optimizing them in complete isolation fails to model the correct visibility and occlusion relationships. To resolve this, we design the \textbf{Composed Gaussian Splatting} strategy. We maintain separate Gaussian sets to apply entity-specific learning rates and densification rules, but critically, we merge them into a unified scene for rasterization. This strategy ensures that: 1) each entity is optimized with appropriate dynamics; and 2) the relative depth and occlusion are correctly learned via the unified 2D projection.

Third, while Composed Gaussian Splatting reconstructs overall visual appearance, it relies primarily on 2D joint human-object mask supervision. Under sparse views, this 2D supervision is often insufficient to constrain the 3D geometry in contact regions, leaving the model blind to the precise physical interface. This leads to visually plausible but physically impossible artifacts, such as inter-penetration or floating contacts. To bridge this gap, we introduce \textbf{Physically Plausible Optimization}, which explicitly injects 3D physically plausible geometric constraints that 2D images cannot provide. Specifically, we design a Contact Loss to eliminate unnatural gaps and a Separation Loss to resolve inter-penetration, serving as the final regularizer to ensure the interaction is not just visually clear, but physically valid.

As illustrated in Figure~\ref{fig:teaser}, HOGS addresses the limitations of existing methods by combining this efficient joint rendering of human and object Gaussians with a physically plausible optimization process. Through this holistic design, HOGS simultaneously achieves high rendering efficiency, superior visual quality, and geometric plausibility—a harmony that previous approaches have not attained.

Our main contributions are summarized as follows:
\begin{itemize}[leftmargin=*]
    \item We present HOGS, a novel framework for sparse-view HOI rendering. Our method achieves a significant performance leap, surpassing state-of-the-art 3DGS-based methods by 3.4 dB in PSNR while maintaining real-time rendering speeds.
    
    \item We propose a novel optimization mechanism for physical plausibility based on geometric consistency, which constitutes the core of our method. Tailored for unstructured 3D Gaussians, this mechanism integrates SDF-based constraints to explicitly enforce non-penetration and accurate contacts.
    
    \item To support this core pipeline under sparse-view ambiguity, we develop a suite of auxiliary components, including a strategy for robust human pose refinement and a contact prediction module for efficient rendering.
    
    \item We demonstrate the effectiveness of HOGS on HOI rendering~\cite{zhang2023neuraldome} and further extend it to hand-object grasp rendering~\cite{pokhariya2024manus}, showcasing its applicability to diverse articulated interactions.
\end{itemize}

%% file: sec/2_related.tex
\section{Related Work}

\noindent\textbf{Free Viewpoint Rendering.}
Free Viewpoint Rendering (FVR) focuses on synthesizing novel views of a scene and has been a long-standing challenge in computer vision~\cite{xian2021space, weng2022humannerf, yang2022neural, jayasundara2023flexnerf, shetty2024holoported}. 
Traditional methods generate novel views by blending or warping input views based on geometric constraints~\cite{smolic2004free, bansal20204d, shum2000review}, but are limited to viewpoints near the input cameras.
Recent research has shifted towards using neural representations, which enable higher-quality FVR both for static scenes and dynamic scenarios~\cite{zhi2021place, wu2022object, cao2023hexplane, ding2024point}. 
However, these methods typically require dense views for higher-quality rendering. 
In real-world applications, dense-view setups are impractical or impossible, leading to sparse-view scenarios due to occlusions or limited camera setups~\cite{xu2024relightable, zhang2025cor, kwon2025generalizable,sun2025real, zhou2023hdhuman}. 
These sparse-view conditions further complicate novel view synthesis, especially for complex HOIs.
To address this challenge, our work focuses on high-quality HOI rendering from sparse views.

\noindent\textbf{Differentiable Rendering of Radiance Fields.}
The advent of differentiable rendering has significantly advanced FVR~\cite{petersen2022gendr, bangaru2022differentiable, worchel2023differentiable}. 
Neural Radiance Fields (NeRF)~\cite{mildenhall2021nerf} pioneered this field by representing scenes as implicit functions~\cite{pumarola2021d, cao2023hexplane, zhou2025nerfect, li2024gp, muller2022instant}. 
While NeRF provides impressive rendering results, its reliance on ray marching remains a computational bottleneck. 
In contrast, 3D Gaussian Splatting (3DGS)~\cite{kerbl20233d} employs explicit 3D Gaussians for fast and efficient rendering, rapidly emerging as a powerful technique for real-time rendering~\cite{zhu2025fsgs, liu2025citygaussian, qian20243dgs}. 
Leveraging its inherent efficiency, 3DGS has been successfully extended to dynamic human modeling, enabling the impressive reconstruction of human avatars~\cite{kocabas2024hugs, moreau2024human, hu2024gauhuman}. 
However, these methods primarily focus on single humans and do not explicitly address the complex interactions inherent in multi-object scenes. 
In this work, we extend 3DGS to jointly model and render HOIs,
enabling efficient and high-quality rendering of complex scenes.

\noindent\textbf{HOI Rendering.}
Early approaches for HOI rendering primarily relied on mesh-based reconstructions, where humans and objects were reconstructed separately and then rendered together~\cite{collet2015high, schonberger2016structure, dou2017motion2fusion}. While straightforward, these methods often struggled with challenges such as occlusions and incomplete textures. More recent research has explored neural rendering techniques for HOIs, including texture blending~\cite{sun2021neural}, volumetric rendering~\cite{zhang2023neuraldome, su2022robustfusion}, and NeRF-based pipelines~\cite{jiang2022neuralhofusion}. Among the latest advancements, NeuralDome~\cite{zhang2023neuraldome} employs a layer-wise neural processing pipeline to render complex interactions. However, it relies heavily on dense multi-view inputs and computationally expensive volumetric rendering, making it less suitable for sparse-view scenarios.

Beyond visual fidelity, enforcing \textbf{\textit{physically plausible}} interactions is crucial for realistic HOI rendering~\cite{yang2021cpf, xie2023visibility}. Traditional methods often incorporated physical optimizations into their mesh-based reconstructions to ensure plausible human-object contact. This was typically achieved using penalty terms~\cite{zhang2024hoi, hassan2019resolving, hassan2021populating}, specific constraints~\cite{hu2024hand, battaglia2016interaction}, or physical criteria~\cite{gupta2009observing, jain2009interactive} that operate on mesh vertices. However, these techniques, designed for static mesh topologies, are not directly applicable to modern explicit representations like 3D Gaussian Splatting. The 3DGS framework relies on a collection of Gaussian primitives that dynamically change in number and position during optimization, rendering mesh-centric constraints ineffective.

To bridge this gap, our work introduces HOGS, which not only leverages 3DGS for an efficient and high-fidelity solution to HOI rendering from sparse views but also pioneers a physically plausible optimization tailored for Gaussian primitives. By using novel Gaussian contact and separation losses, our method is the first to enforce physically plausible HOIs directly within the 3DGS framework, significantly enhancing both realism and rendering quality.

%% file: sec/3_method.tex
\section{Methodology}
\label{method}

\begin{figure*}[t]
\begin{center}
\includegraphics[width=1.0\linewidth]{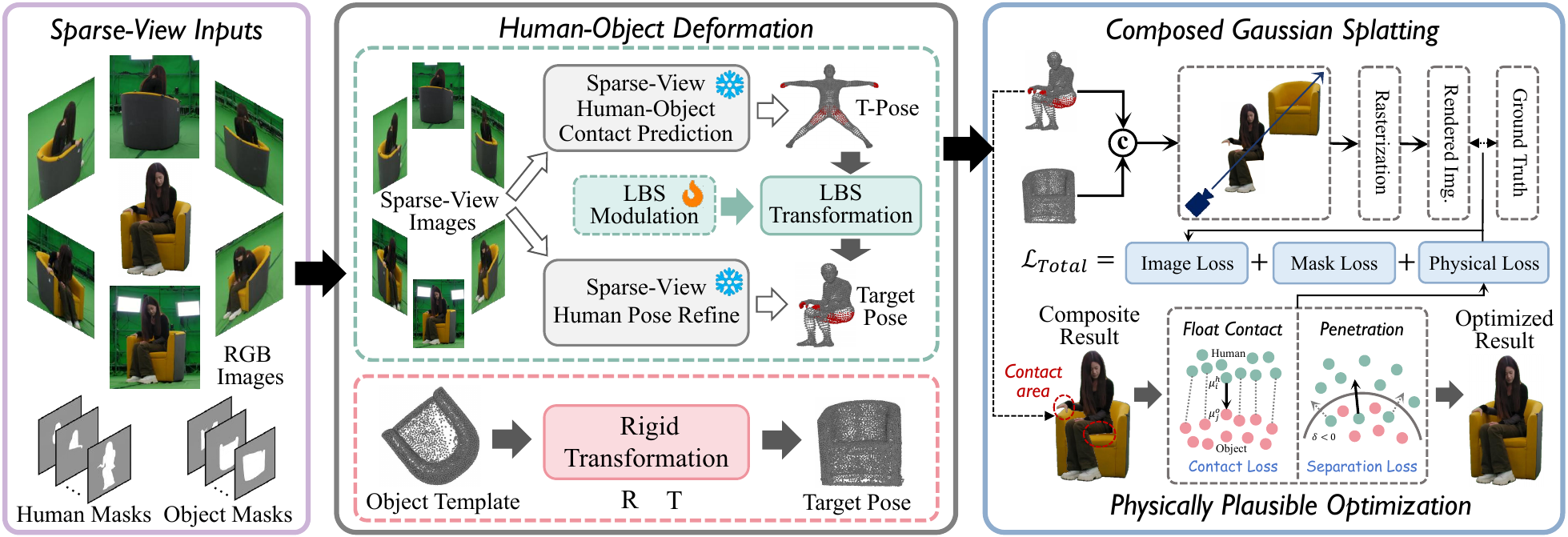}
\end{center}
\caption{\textbf{Overview of the HOGS pipeline.} Our framework takes sparse-view images of an HOI scene as input and consists of three main stages.
\textbf{(1) \emph{Human-Object Deformation}:} The canonical human and object models are deformed to match the current scene's pose. This process is driven by two key pre-trained modules processing the sparse inputs: the \texttt{Sparse-View Human Pose Refiner} (frozen \frozenicon) aggregates sparse-view information to predict robust pose parameters, which then drive the \texttt{LBS Transformation}. Simultaneously, the learnable \texttt{LBS Modulation} (trainable \trainableicon) refines the skinning weights to capture non-rigid details. The object template is rigidly transformed to match the target pose.
\textbf{(2) \emph{Composed Gaussian Splatting}:} The posed models are converted into distinct sets of human and object Gaussians, which are then merged into a unified representation for joint, efficient rendering.
\textbf{(3) \emph{Physically Plausible Optimization}:} Finally, the rendering is refined to ensure physical plausibility. This stage utilizes the contact mask predicted by the \texttt{Sparse-View Human-Object Contact Prediction} module (frozen \frozenicon) to focus the computation of our contact and separation losses on relevant interaction regions.}
\label{fig:Overview}
\end{figure*}

For each frame of a dynamic HOI scene, our HOGS pipeline takes as input $N$ sparse-view RGB images $\{I_i\}_{i=1}^N$, along with their corresponding human-object foreground masks $\{M^{ho}_i\}_{i=1}^N$.
The pipeline operates on a hybrid optimization strategy (as shown in Figure~2):
First, to overcome the ambiguity of sparse views, we leverage two powerful, generalizable modules (marked with \frozenicon)---the \emph{Sparse-View Human Pose Refinement} module and the \emph{Sparse-View Human-Object Contact Prediction} module. These networks are pre-trained only once and function as robust off-the-shelf estimators to provide initial pose parameters and contact regions.
Second, these estimations serve as the foundation for our per-scene optimization, where we employ a learnable \emph{LBS Modulation} (marked with \trainableicon) and optimize the Gaussian attributes to recover fine-grained details.
This strategy separates the costly training of generalizable models from the efficient, per-scene optimization of the final Gaussian representation. To provide a clear roadmap of this process, the step-by-step training procedure is formally summarized in Algorithm~\ref{alg:hogs_pipeline}.
The HOGS pipeline comprises three main stages, corresponding to the step-by-step data flow visualized in Figure~2 and detailed sequentially in Sec.~III-A (Human-Object Deformation), Sec.~III-B (Composed Gaussian Splatting), and
Sec.~III-C (Physically Plausible Optimization).

\subsection{Human-Object Deformation}
\label{sec:human_object_deformation}

To accurately represent HOIs, we first deform both the human and the object from initial states to their respective target states. 
Human deformation includes applying standard LBS transformations, adding an LBS modulation to capture finer details, and further refining the target pose through a refinement module. 
The object deformation is simplified by treating the object as rigid and estimating its rigid transformation with respect to a template mesh.

\subsubsection{LBS Transformation}
LBS is commonly employed in human rendering to deform human representations~\cite{liu2024humangaussian, kocabas2024hugs, hu2024gauhuman, moreau2024human}. 
Following this approach, we utilize LBS to transform human Gaussians. 
We adopt the SMPL-H~\cite{romero2022embodied} model, a parametric 3D human body model that extends SMPL to incorporate high-fidelity hand details. 
3D Gaussians are initialized by placing their means $\mathbf{p}^c$ at the SMPL-H mesh vertices and assigning each an initial covariance $\mathbf{\Sigma}^c$. 
These initial Gaussians, defined in the canonical T-pose space, are then transformed to the posed space via LBS.  
Specifically, the transformed mean $\mathbf{p}^t$ and covariance $\mathbf{\Sigma}^t$ are given by:
\begin{gather}
    \mathbf{p}^t = \textstyle\sum_{k=1}^{K} w_k (\mathbf{R}_k \mathbf{p}^c + \mathbf{t}_k) + \mathbf{b}, \\
    \mathbf{\Sigma}^t = \left(\textstyle\sum_{k=1}^{K} w_k \mathbf{R}_k\right) \mathbf{\Sigma}^c \left(\textstyle\sum_{k=1}^{K} w_k \mathbf{R}_k\right)^T,
\end{gather}
where $K$ is the number of joints, $w_k$ is the LBS weight associated with joint $k$, $\mathbf{b}$ is a global translation vector, and $\mathbf{R}_k$ and $\mathbf{t}_k$ represent the rotation matrix and translation vector.

\subsubsection{LBS Modulation}
While LBS provides an efficient way to deform the human representation, it is a linear transformation strictly limited to the underlying mesh topology, 
Consequently, it often struggles to capture fine-grained details and subtle deformations.
Inspired by previous works that learn LBS weight fields for detailed human modeling~\cite{huang2020arch, peng2021animatable, hu2024gauhuman, lin2022learning, liu2023hosnerf}, we introduce an LBS modulation to refine the initial LBS weights derived from the SMPL-H model.
Specifically, we leverage the pre-computed SMPL-H weights as a strong prior and employ an MLP $\Phi_{lbs}$ to modulate each LBS weight. 
Given a 3D Gaussian centered at $\mathbf{p}^c$ in the canonical space, we first apply a positional encoding $\gamma(\mathbf{p}^c)$ to its position. 
The MLP then outputs a modulation vector $\mathbf{m} = \Phi_{lbs}(\gamma(\mathbf{p}^c))$, where each element $m_k$ corresponds to the modulation factor for the $k$-th LBS weight. 
The final LBS weight $w_k$ is computed through a softmax function:
\begin{equation}
    w_k = \texttt{softmax}(w_k^\text{SMPL-H} + m_k)
\end{equation}
where $w_k^{\text{SMPL-H}}$ is the LBS weight derived from the nearest SMPL-H vertex for joint $k$. 
This approach modulates LBS weights efficiently, capturing finer details of deformations.

\subsubsection{Sparse-View Human Pose Refinement}
\begin{figure}[t]
\centering
\includegraphics[width=0.49\textwidth]{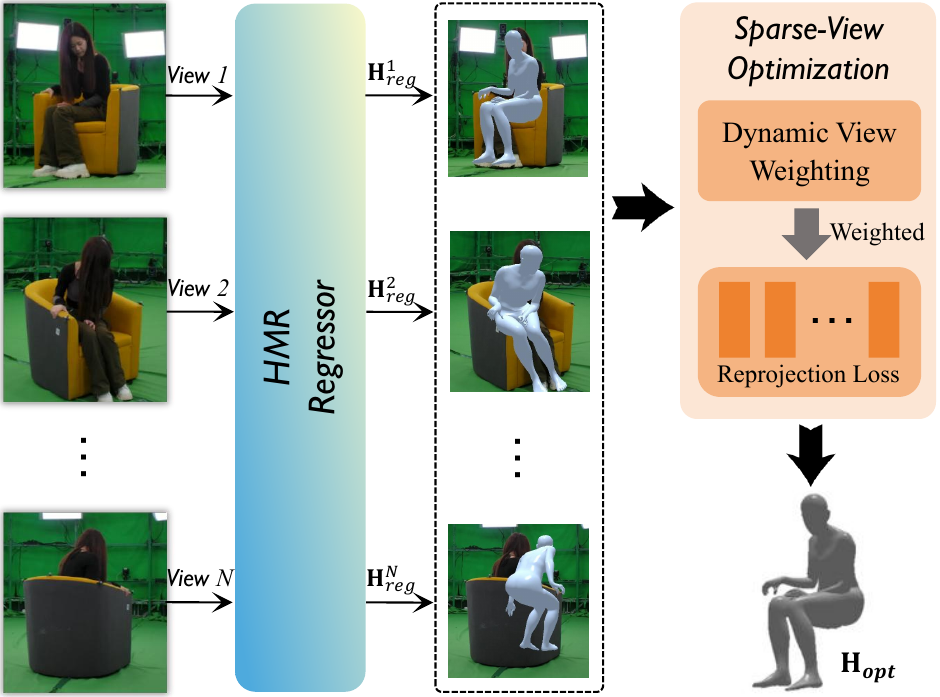} % 
\caption{\textbf{Training strategy for the Sparse-View Human Pose Refinement module.} 
We employ an optimization-in-the-loop approach to train our HMR-based regressor. For each training sample, the regressor's initial per-view estimates are refined by a differentiable multi-view optimization process, which leverages dynamic view weights to handle occlusions. The regressor is then trained end-to-end to directly predict this refined, multi-view consistent pose.}
\label{fig:human pose refine}
\end{figure}

Accurate LBS transformations depend on a reliable target human pose, but achieving this from sparse views is challenging due to frequent human-object occlusions. 
Existing multi-view optimization methods typically rely on the triangulation of 2D keypoints or latent feature fusion~\cite{Qiu_2019_ICCV, he2020epipolar, iskakov2019learnable, ma2021transfusion, wan2023view, zhang2021adafuse}. 
However, triangulation degrades severely when 2D detections are noisy due to occlusion, while feature fusion struggles to generalize to sparse camera setups without extensive pre-training.
To overcome these limitations, we develop a Sparse-View Human Pose Refinement module.
Unlike methods that implicitly learn fusion weights, we introduce a \textbf{\emph{Dynamic View Weighting}} mechanism to explicitly filter out occluded noise based on physical occlusion rates.
This ensures that the optimization is strictly driven by reliable visual cues.
Furthermore, rather than performing costly test-time optimization for each new scene, our goal is to build a powerful, generalizable regressor that can directly infer high-quality poses in a single forward pass.
As illustrated in Figure~\ref{fig:human pose refine}, this strategy leverages a multi-view optimization process during the training phase to teach a standard regressor how to handle occlusions and fuse information from sparse views.

\noindent\textbf{Training Strategy.}
Our training strategy revolves around an HMR-based regressor~\cite{kanazawa2018end} and a novel, optimization-based training objective. For each training sample, the regressor first produces an initial set of SMPL-H parameters $\mathrm{H}^{i}_{\text{reg}} = \{\theta^{i}_{\text{reg}}, \beta^{i}_{\text{reg}}\}$ for each of the $N$ sparse views. Instead of using these predictions directly in a simple loss, we use them to initialize a differentiable optimization process that finds a more physically plausible, multi-view consistent pose, which we denote as $\mathrm{H}_{\text{opt}}$. The final loss function then encourages the regressor's direct output, $\mathrm{H}^{i}_{\text{reg}}$, to be as close as possible to the target pose $\mathrm{H}_{\text{opt}}$.

The core of this strategy lies in defining the optimization objective for finding $\mathrm{H}_{\text{opt}}$. This process is inspired by SMPLify~\cite{bogo2016keep} and is adapted for our sparse-view setting by aggregating per-view costs. The cost for the $i$-th view is:
\begin{equation}
\mathcal{E}_{i} = \lVert P(\mathrm{H}) - J^{i}_{\text{reg}} \rVert^2 + \lambda_\theta E_\theta(\theta) + \lambda_\beta E_\beta(\beta),
\end{equation}
where $P(\mathrm{H})$ is the 2D projection of the estimated 3D joints, $J^{i}_{\text{reg}}$ are the initial 2D joints predicted by the regressor, and $E_\theta(\theta), E_\beta(\beta)$ are regularization terms~\cite{li2019parametric} with weights $\lambda_{\theta}=1$ and $\lambda_{\beta}=0.001$.

\noindent\textbf{Dynamic View Weighting.} To handle the varying reliability of different views due to occlusions, we introduce dynamic view weights $d_i$. We estimate an occlusion rate $O_{i} = 1 - V_i / K$, where $V_i$ is the number of visible joints within the human mask in view $i$ and $K$ is the total joint count. These rates are used to compute the weights via $d_i = \texttt{softmax}(-\alpha O_{i})$, where $\alpha=5$ is a sensitivity factor. The total optimization cost is then the weighted sum $\mathcal{E}_{\text{total}} = \sum_{i=1}^{N} d_i \cdot \mathcal{E}_{i}$. By minimizing this cost, we obtain the optimized target pose $\mathrm{H}_{\text{opt}} = \{\theta_{\text{opt}}, \beta_{\text{opt}}\}$. It is crucial to note that this optimization is a differentiable operation \textbf{\emph{within the training loop}}, serving to produce a high-quality supervision signal for our regressor.

\noindent\textbf{End-to-End Regressor Training.}
The HMR-based regressor is trained end-to-end to minimize a composite loss function that leverages the optimized pose target $\mathrm{H}_{\text{opt}}$:
\begin{equation}
\mathcal{L}_{\text{HPR}} = \lambda_1 \mathcal{L}_{2D} + \lambda_2 \mathcal{L}_{3D} + \lambda_3 \mathcal{L}_{\text{H}},
\end{equation}
where $\mathcal{L}_{2D}=\sum_{i=1}^N \lVert J^{i}_{\text{reg}} - J^{i}_{\text{gt}} \rVert$ is a standard 2D reprojection loss. The 3D loss $\mathcal{L}_{3D}=\sum_{i=1}^N \lVert \mathbf{p}^{i}_{\text{opt}} - \mathbf{p}^{i}_{\text{gt}} \rVert$ supervises the optimized 3D joints $\mathbf{p}^{i}_{\text{opt}}$ against the ground truth. Most importantly, the consistency loss $\mathcal{L}_{\text{H}}=\sum_{i=1}^N \lVert \mathrm{H}^{i}_{\text{reg}} - \mathrm{H}_{\text{opt}} \rVert$ explicitly forces the regressor's direct output to match the pose found by the optimization process. In this way, the network learns to internalize the logic of multi-view fusion and occlusion handling. The loss weights are set to $\lambda_1=5$, $\lambda_2=5$, and $\lambda_3=0.001$.

\noindent\textbf{Inference on New Scenes.}
The described optimization-guided training strategy yields a pose refinement module that is both powerful and efficient. 
At inference time, for any new or unseen scene, we perform a \textbf{\emph{single forward pass}} through the trained regressor to obtain the refined human pose parameters. 
Crucially, no per-scene optimization is required, which enables robust and efficient pose estimation.

\subsubsection{Object Deformation}
Objects involved in HOIs are typically rigid. 
Following prior work~\cite{taheri2020grab, zhang2023neuraldome}, we estimate object rotation $R_t \in SO(3)$ and translation $T_t \in \mathbb{R}^3$ relative to a template mesh $\mathcal{M}_{\text{temp}}$. 
As shown in Fig.~\ref{fig:Overview}, we employ ICP~\cite{besl1992method} to register $\mathcal{M}_{\text{temp}}$ to per-frame 3D object markers. 
These markers ensure robust ICP initialization, mitigating local minima issues and yielding accurate per-frame poses ($R_t, T_t$) for the target object mesh $\mathcal{M}_{\text{tar}}$. 
This enables tracking the rigid motion of objects.
It is worth noting that we strategically employ this stable and established rigid solver to ensure system stability. This design choice allows us to allocate computational resources to our core innovation---the Physically Plausible Optimization---which resolves the more critical and challenging issue of human-object interaction.

\subsection{Composed Gaussian Splatting}
\label{sec:composed_gaussian_splatting}

We extend 3DGS to jointly render humans and objects. While treating the scene as a single Gaussian set might seem straightforward, it leads to \emph{\textbf{structural conflicts}} due to the disparate characteristics of the entities: the high-frequency motion gradients from the articulated human often trigger erroneous densification in the adjacent rigid object.
To resolve this, our strategy adheres to two design principles: (1) \textbf{\emph{Decoupled Learning Dynamics}}, where human and object Gaussians are maintained as distinct subsets to preserve their specific structural properties (i.e., articulated vs. rigid); and (2) \textbf{\emph{Unified Differentiable Sorting}}, where these subsets are merged prior to rasterization. This ensures that despite being maintained separately, they enter the same visibility sorting process, which is crucial for resolving depth ambiguities and occlusion relationships in sparse-view settings.

Based on this design, we model the scene with two distinct Gaussian subsets: 
1) \textbf{\emph{Human Gaussians}}: For the human, each vertex of the SMPL-H model is converted into a 3D Gaussian, forming the human Gaussian set $\{\mathcal{G}^h_i = \mathcal{N}(\mathbf{x}_i; \mu_i^{h}, \Sigma_i^{h})\}$, where $\mathcal{N}(\mathbf{x}; \mu, \Sigma)$ represents a Gaussian distribution with mean $\mu$ and covariance $\Sigma$.
The human Gaussian parameters $\mu_i^{h}$ and $\Sigma_i^{h}$ are updated frame-by-frame based on the refined pose and shape parameters from Sec.~\ref{sec:human_object_deformation}, ensuring accurate motion tracking of the articulated human body. 
2) \textbf{\emph{Object Gaussians}}: For the object, the vertices of the target pose of object mesh $\mathcal{M}_{\text{tar}}$ are used to initialize the means of the object Gaussians. 
Specifically, each vertex of $\mathcal{M}_{\text{tar}}$ becomes the mean $\mu_j^o$ of an object Gaussian $\mathcal{G}^o_j$, with an assigned initial covariance $\Sigma_j^o$. 
Finally, the human and object Gaussian sets are merged into a unified set: $\mathcal{G} = \{\mathcal{G}^h_i\} \cup \{\mathcal{G}^o_j\}$.

The combined Gaussian set $\mathcal{G}$ is projected into the 2D plane.
The covariance matrix $\Sigma$ of each 3D Gaussian is transformed to its projected 2D covariance $\Sigma'$ as follows:
\begin{equation}
    \Sigma' = \mathbf{J} \, \mathbf{W} \, \Sigma \, \mathbf{W}^{T} \mathbf{J}^{T},
\end{equation}
where $\mathbf{W}$ is the world-to-camera transformation matrix, $\mathbf{J}$ is the Jacobian of the affine approximation of the projective transformation.
The projected Gaussians are blended using $\alpha$-compositing to generate the rendered image, which is compared with the ground-truth image using $\mathcal{L}_{\text{image}}$ loss.

\subsection{Physically Plausible Optimization}
\label{sec:physics_aware_optimization}

To ensure physically plausible HOI rendering, we introduce a two-step rendering optimization process. 
Firstly, we predict potential human-object contact regions using a novel sparse-view human-object contact prediction module. 
This prediction yields a set of human Gaussians corresponding to the contact regions. 
Secondly, leveraging these predicted regions, we perform a physical optimization on the composed Gaussian splatting result to enforce physical constraints.

\subsubsection{Sparse-View Human-Object Contact Prediction}
\label{sec:sparse_view_contact_prediction}

\begin{figure}[t]
\centering
\includegraphics[width=0.45\textwidth]{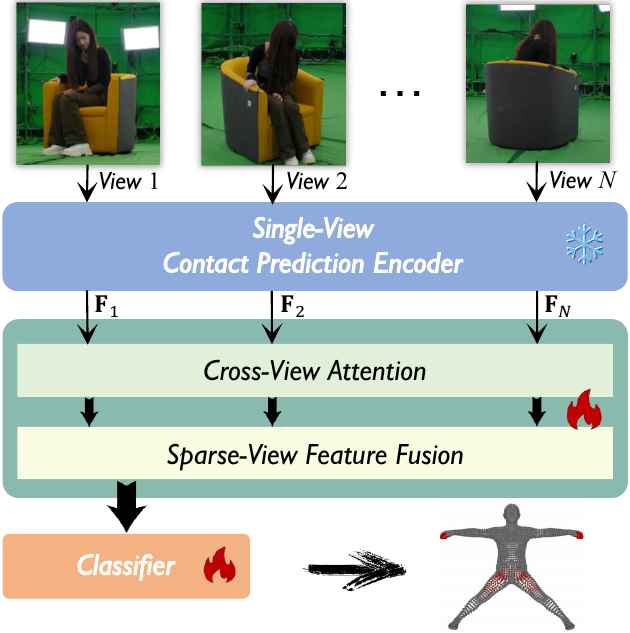} % 
\caption{\textbf{Architecture of the Sparse-View Contact Prediction module.} 
Given sparse-view images, features are first extracted from each sparse input view using a shared encoder. A \textbf{\emph{cross-view attention}} module then fuses these features to aggregate multi-view information. Finally, a classifier predicts a per-vertex contact probability map on the human mesh to identify regions of interaction.}
\label{fig:sparse_view_contact_prediction}
\end{figure}

\begin{algorithm}[t]
\caption{Overview of HOGS Pipeline}
\label{alg:hogs_pipeline}
\small 
\SetAlgoLined
\DontPrintSemicolon
\KwIn{Sparse images $\{I_i\}_{i=1}^N$, Masks $\{M^{ho}_i\}_{i=1}^N$, Pre-trained Modules (Refiner, Contact Predictor).}
\KwOut{Optimized Human-Object Gaussians $\mathcal{G}$.}

\tcp{--- Initialization \& Pre-computation ---}
Estimate human pose $\mathrm{H}_{\text{opt}}$ via Sparse-View Pose Refiner;\;
Estimate object pose $(R_t, T_t)$ and obtain $\mathcal{M}_{\text{tar}}$;\;
Compute SDF of $\mathcal{M}_{\text{tar}}$ for $\mathcal{L}_{\text{sep}}$ calculation;\;
Predict contact vertex set $\mathcal{C}$ via Contact Predictor;\;
Initialize human Gaussians $\mathcal{G}^h$ and object Gaussians $\mathcal{G}^o$;\;

\tcp{--- Training Pipeline ---}
\While{not converged}{
    \tcp{Human-Object Deformation}
    Compute LBS modulation $\mathbf{m} = \Phi_{lbs}(\gamma(\mathbf{p}^c))$;\;
    Refine $w_k$ and transform $\mathcal{G}^h$ to $\mathbf{p}^t$ via LBS (Eq.~(1)--(3));\;
    
    \tcp{Composed Gaussian Splatting}
    Compose unified scene: $\mathcal{G} \leftarrow \{\mathcal{G}^h\} \cup \{\mathcal{G}^o\}$;\;
    Project $\mathcal{G}$ to 2D views and rasterize to rendered images $\hat{I}$;\;
    
    \tcp{Physically Plausible Optimization}
    Compute rendering losses: $\mathcal{L}_{\text{image}}, \mathcal{L}_{\text{ssim}}, \mathcal{L}_{\text{lpips}}, \mathcal{L}_{\text{mask}}$;\;
    Compute $\mathcal{L}_{\text{contact}}$ on set $\mathcal{C}$ (Eq.~7);\;
    Compute $\mathcal{L}_{\text{sep}}$ using pre-computed SDF (Eq.~8);\;
    $\mathcal{L}_{\text{total}} \leftarrow$ Sum of weighted losses (Eq.~9);\;
    Update Gaussian parameters and LBS MLP $\Phi_{lbs}$;\;
}
\Return Final $\mathcal{G}$
\end{algorithm}

Directly compositing Gaussians (as in Sec.~\ref{sec:composed_gaussian_splatting}) may lead to physically implausible interactions at human-object contact regions.
However, enforcing physical constraints globally is computationally prohibitive, as it would necessitate calculating signed distance fields (SDF) for all dynamic Gaussians at every iteration.
To ensure optimization feasibility, our strategy is to first efficiently identify the specific human-object contact regions. 
By doing so, we focus the subsequent physical optimization only on the relevant subset of primitives, avoiding expensive global queries.
To this end, we propose a sparse-view contact prediction module, which is illustrated in Fig.~\ref{fig:sparse_view_contact_prediction}.

\noindent\textbf{Architecture.}
Our module first extracts per-view HOI features from the $N$ input views using a single-view contact prediction encoder. Specifically, we adapt the DECO~\cite{tripathi2023deco} architecture, excluding its final MLP layer, to produce per-view features $\mathrm{F} = \{\mathrm{F}_i\}_{i=1}^N \in \mathbb{R}^{N \times D}$, where the feature dimension $D=2048$. These features $\mathrm{F}$ are then processed by a cross-view attention module to aggregate information across all views. Within the attention module, the query, key, and value matrices ($\mathbf{Q}$, $\mathbf{K}$, $\mathbf{V}$) are computed by linearly projecting the input features $\mathrm{F}$ using weight matrices $W_Q, W_K, W_V \in \mathbb{R}^{2048 \times 512}$. The attention mechanism outputs a tensor $\mathrm{F}_{\text{att}} \in \mathbb{R}^{N \times 512}$. Subsequently, a sparse-view feature fusion module averages $\mathrm{F}_{\text{att}}$ across the view dimension to produce a single fused feature vector $\mathrm{F}_{\text{fuse}} \in \mathbb{R}^{512}$. Finally, a lightweight classifier takes $\mathrm{F}_{\text{fuse}}$ as input to yield contact probabilities $\mathcal{P} \in \mathbb{R}^{6890}$ for all body vertices of the SMPL-H model.

\noindent\textbf{Training Process.}
The module is trained in a supervised manner. 
We adopt a transfer learning approach by initializing the single-view encoder with weights from the pre-trained DECO model~\cite{tripathi2023deco}. 
During our training phase, the parameters of this encoder are frozen to retain its powerful feature extraction capabilities. 
We then train only the newly added cross-view attention module, the feature fusion module, and the final classifier. 
The training objective is a standard binary cross-entropy loss between the predicted probabilities $\mathcal{P}$ and the ground truth labels. 

\noindent\textbf{Inference Stage.}
At inference time, the pre-trained contact prediction module takes the $N$ sparse views of a new scene as input. 
It performs a single forward pass to directly output the contact probabilities $\mathcal{P}$. 
These probabilities are then thresholded (with $\tau=0.5$) to identify the contact vertex set $\mathcal{C}$, which is subsequently used to guide the physically plausible optimization (Sec.~\ref{sec:physical_optimization}).

\subsubsection{Physically Plausible Optimization}
\label{sec:physical_optimization}

In this context, we focus on establishing geometric consistency by specifically enforcing constraints such as non-penetration and surface contact. 
To enforce these constraints, previous mesh-based methods~\cite{bhatnagar2022behave, xie2022chore, jiang2023full, dabral2021gravity} typically rely on vertex-to-vertex distances calculated via Nearest Neighbor (NN) search.
However, directly applying such mesh-centric losses to 3DGS is computationally prohibitive. Unlike meshes with fixed semantic vertices ($\sim$6K), 3DGS involves massive, unstructured primitives ($\sim$10k-100K) that change dynamically. An explicit NN search for every Gaussian would result in intractable complexity ($O(N \log M)$).
To address this, we propose a scalable alternative: we replace explicit neighbor searching with efficient Signed Distance Field (SDF) queries ($O(1)$ per Gaussian) and utilize our predicted contact mask to filter optimization targets. This combination allows us to enforce physical plausibility with negligible computational overhead.

With human Gaussians in contact regions identified (yielding the vertex index set $\mathcal{C}$ from contact prediction, Sec.~\ref{sec:sparse_view_contact_prediction}), we perform physically plausible optimization on these regions within our HOGS framework (Figure~\ref{fig:Overview}). 
This optimization enforces physical plausibility by minimizing our \textit{Gaussian Contact and Separation Losses}, guiding human and object Gaussians towards plausible configurations.

\noindent\textbf{Contact Loss.}
To encourage closer human-object Gaussian proximity in regions $\mathcal{C}$, we define the contact loss. 
Specifically, this loss considers distances between human Gaussians $\{\mathcal{G}^h_i(\mu_i^h, \Sigma_i^h)\}_{i \in \mathcal{C}}$ (for contact vertices in $\mathcal{C}$) and all $N_o$ object Gaussians $\{\mathcal{G}^o_j(\mu_j^o, \Sigma_j^o)\}_{j=1}^{N_o}$, which is defined as:
\begin{equation}
\small
    \mathcal{L}_{\text{contact}} = \frac{1}{|\mathcal{C}|} \sum_{i \in \mathcal{C}} \min_{1 \le j \le N_o} \| \mu_i^h - \mu_j^o \|_2 
    + \frac{1}{N_o} \sum_{j=1}^{N_o} \min_{i \in \mathcal{C}} \| \mu_j^o - \mu_i^h \|_2.
\label{eq:contact_loss} 
\end{equation}
This loss term effectively pulls the interacting regions of the human and object Gaussians closer, promoting a more physically plausible interaction by reducing unnatural gaps.

\noindent\textbf{Separation Loss.}
Besides reducing unnatural gaps via the contact loss, intuitive physics dictates that objects cannot occupy the same 3D space, a phenomenon known as \emph{penetration}. 
Thus, we introduce a separation loss $\mathcal{L}_{\text{sep}}$ to penalize human-object Gaussian interpenetration. 
Our approach leverages Signed Distance Fields (SDFs) for efficient penetration detection. 
Directly computing SDFs from all dynamic Gaussians incurs a significant computational burden, as SDFs would need frequent recomputation during optimization due to changes in Gaussian numbers and positions. 
To mitigate this, we exploit the static nature of the object, represented by a fixed target mesh $\mathcal{M}_{\text{tar}}$ (Sec.~\ref{sec:human_object_deformation}). 
Consequently, we pre-compute the SDF of $\mathcal{M}_{\text{tar}}$ only once before 3DGS optimization.

We construct a uniform $256 \times 256 \times 256$ voxel grid over a padded bounding box of $\mathcal{M}_{\text{tar}}$. 
Each voxel center $c_t \in \mathbb{R}^3$ stores its signed distance $\delta_t$ to the nearest surface point on $\mathcal{M}_{\text{tar}}$, where the sign of $\delta_t$ indicates if $c_t$ is inside (negative) or outside (positive) the object. 
Penetration of a human Gaussian $\mathcal{G}^h_i$ is determined by querying the SDF at its mean $\mu_i^h$. 
Trilinear interpolation within the grid yields a continuous signed distance $\delta(\mu_i^h)$ and normal $\mathbf{n}(\mu_i^h)$ at $\mu_i^h$. 
The separation loss is then defined as:
\begin{equation}
    \mathcal{L}_{\text{sep}} = \frac{1}{|\mathcal{C}|} \sum_{i \in \mathcal{C}} \max(0, -\delta(\mu_i^h)) \| \mathbf{n}(\mu_i^h) \|_2^2,
\label{eq:separation_loss} 
\end{equation}
where we only penalize negative $\delta(\mu_i^h)$ values (points inside the object).

\noindent\textbf{Total Loss.}
As shown in Figure~\ref{fig:Overview}, the total training loss for composed Gaussian splatting is:
\begin{equation}
% \small
\begin{aligned}
    \mathcal{L}_{\text{total}} = & \mathcal{L}_{\text{image}} + \lambda_{\text{ssim}} \mathcal{L}_{\text{ssim}} + \lambda_{\text{lpips}} \mathcal{L}_{\text{lpips}} \\ & 
    + \lambda_{\text{mask}} \mathcal{L}_{\text{mask}}
    + \lambda_{\text{contact}} \mathcal{L}_{\text{contact}} + \lambda_{\text{sep}} \mathcal{L}_{\text{sep}},
\end{aligned}
\end{equation}
where $\mathcal{L}_{\text{mask}}$ uses human-object foreground masks $\{M^{ho}_i\}_{i=1}^N$, while $\mathcal{L}_{\text{image}}$ (Sec.~\ref{sec:composed_gaussian_splatting}), SSIM loss~\cite{wang2004image}, and LPIPS loss~\cite{zhang2018unreasonable} ensure image fidelity against ground truth.

%% file: sec/4_exp.tex
\section{Experiments}
\label{sec:Experiments}

\subsection{Experimental Settings and Implementation Details
}
\label{exp_set}

\begin{figure*}[t]
    \begin{center}
    \includegraphics[width=1.0\linewidth]{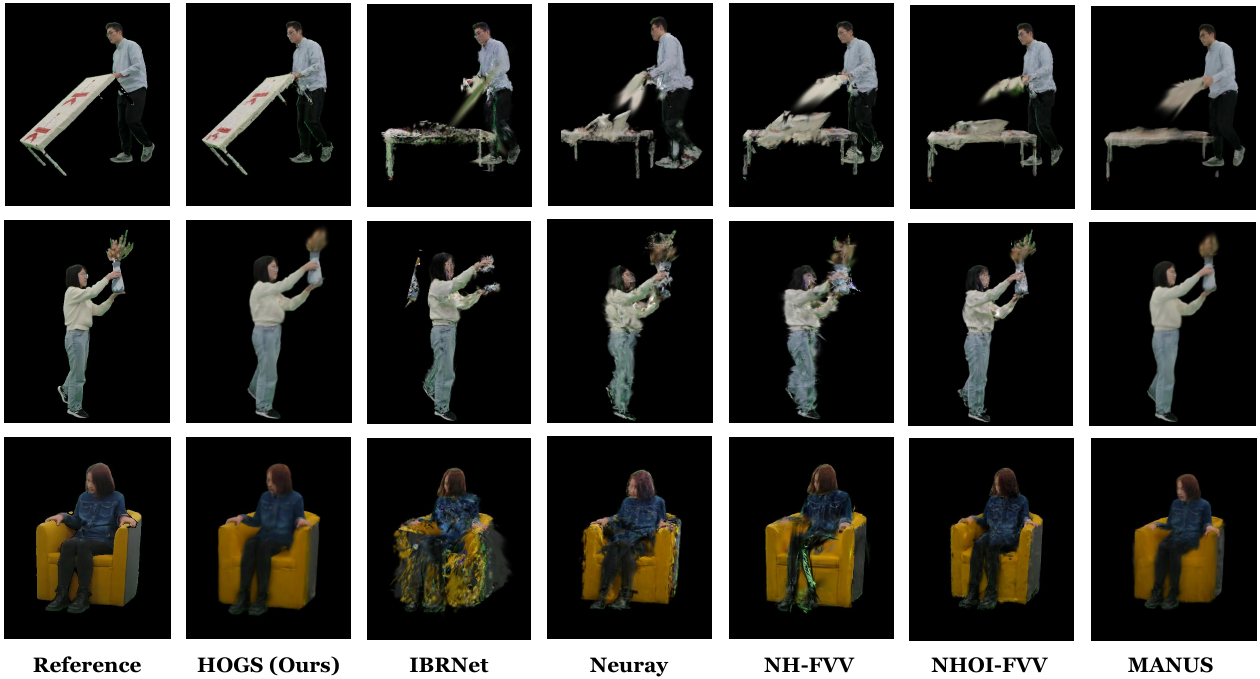}
    \end{center}
    \caption{\textbf{Qualitative comparison on the HODome dataset.} 
    HOGS renders novel views with superior visual fidelity and more plausible HOIs compared to existing methods. In contrast, NeRF-based approaches (\textit{e.g.}, IBRNet, NeuRay) suffer from significant artifacts like blurriness and geometric distortions. The mesh-based NH-FVV exhibits incomplete geometry, while the adapted 3DGS-based MANUS, although structurally sound, lacks fine-grained detail, especially in contact regions.}
    \label{fig:qualitative}
    \vspace{-1em}
\end{figure*}

\subsubsection{Datasets} 
We conduct experiments on two datasets: the large-scale HOI rendering dataset \textbf{HODome}~\cite{zhang2023neuraldome} and the hand-object grasp rendering dataset \textbf{MANUS-Grasps}~\cite{pokhariya2024manus}.
\textbf{HODome} features 76 dynamic HOI sequences with 23 objects and 10 subjects. The dataset provides multi-view RGB videos, human/object masks, object templates, and initial SMPL-H parameters. 
For our experiments, we follow a sparse-view protocol, using 6 fixed input views to reconstruct each interaction scene.
\textbf{MANUS-Grasps} is used to demonstrate HOGS's applicability to more fine-grained articulated interactions. 
It contains high-fidelity hand-object grasp sequences captured by 53 cameras. 
Since this dataset does not provide pre-scanned object templates, we showcase the flexibility of our method by directly using its well-trained object Gaussians, bypassing our object deformation module. 
Similarly, for MANUS-Grasps, we adopt the same sparse-view protocol following~\cite{zhang2023neuraldome}, utilizing 6 fixed input views for reconstruction.

\begin{table}[t]
\centering
\caption{\textbf{Quantitative results of novel view synthesis on HODome dataset.} NH-FVV denotes NeuralHumanFVV, and NHOI-FVV denotes NeuralHOIFVV. We use \colorbox{red!20}{red} and \colorbox{yellow!40}{yellow} text to denote the best and second-best results of each metric respectively.}
    \centering
    \setlength{\tabcolsep}{3pt}
    \renewcommand{\arraystretch}{1.2}
    \small
    \begin{tabular}{cccccc}
        \toprule
        Category & Method & PSNR $\uparrow$ & SSIM $\uparrow$ & LPIPS $\downarrow$ & FPS $\uparrow$ \\
        \midrule
        Mesh-based & NH-FVV & 21.69 & 0.914 & - & 12 \\
        \midrule
        \multirow{3}{*}{NeRF-based} & IBRNet & 21.43 & 0.892 & - & - \\
        & NeuRay & 23.34 & 0.909 & - & 2 \\
        & NHOI-FVV & 23.10 & 0.912 & - & 1 \\
        \midrule
        \multirow{2}{*}{3DGS-based} & MANUS & \cellcolor{yellow!40}27.28 & \cellcolor{yellow!40}0.936 & \cellcolor{yellow!40}0.062 & \cellcolor{yellow!40}155 \\
        & HOGS (Ours) & \cellcolor{red!20}30.68 & \cellcolor{red!20}0.953 & \cellcolor{red!20}0.028 & \cellcolor{red!20}162 \\
        \bottomrule
    \end{tabular}
\label{tab:quantitative}
\end{table}

\subsubsection{Evaluation Metrics}
To quantitatively evaluate the quality of rendered novel view and novel pose images, we evaluate the rendering quality using standard metrics following NeuralDome~\cite{zhang2023neuraldome}: Peak Signal-to-Noise Ratio (PSNR), Structural Similarity Index Measure (SSIM)~\cite{wang2004image}, and Learned Perceptual Image Patch Similarity (LPIPS)~\cite{zhang2018unreasonable}.

\subsubsection{Pre-training of Generalizable Modules}
As detailed in Sec.~\ref{method}, our pose refinement and contact prediction modules are pre-trained to be generalizable. 
We pre-train both modules on the HODome dataset. 
To rigorously evaluate generalization to unseen individuals, we partition the 9 available subjects into distinct training, validation, and testing sets.
Specifically, our training set comprises 6 subjects (\texttt{01, 02, 03, 04, 06, 07}). 
The validation set contains subject \texttt{08}, used for hyperparameter tuning and model selection. 
The test set consists of 2 completely unseen subjects (\texttt{09, 10}) for performance reporting.
For the \textit{Sparse-View Human Pose Refinement} module, we use the Adam optimizer with a learning rate of $3 \times 10^{-5}$ and train for 20 epochs.
For the \textit{Sparse-View Human-Object Contact Prediction} module, we use the Adam optimizer with a learning rate of $1 \times 10^{-5}$ and train for 100 epochs.

\subsubsection{HOGS Implementation Details}
The following details pertain to the per-scene optimization of the main HOGS pipeline. 
All experiments are conducted on a single NVIDIA A100 GPU.
The \textit{LBS Modulation} module (Sec.~\ref{sec:human_object_deformation}) is an MLP with three hidden layers and ReLU activations. It takes a 63-dimensional positional encoding of the canonical Gaussian means as input and outputs a 52-dimensional modulation vector. This MLP is optimized with the rest of the framework using the Adam optimizer with a learning rate of $1 \times 10^{-5}$.
For the \textit{Composed Gaussian Splatting} stage, other 3DGS-related training details (e.g., initialization, densification, pruning, optimization schedule) follow the original implementation~\cite{kerbl20233d}.
The weights for the total loss function (Sec.~\ref{sec:physical_optimization}) are set as follows: $\lambda_{\text{mask}}=0.3$, $\lambda_{\text{ssim}}=0.5$, $\lambda_{\text{lpips}}=0.1$, $\lambda_{\text{contact}}=0.01$, and $\lambda_{\text{sep}}=0.01$. These hyperparameters, particularly the physical loss weights, are determined through a sensitivity analysis on the validation set (Subject 08) to achieve the optimal balance between geometric plausibility and visual fidelity (see Sec.~\ref{exp:ablation} for detailed analysis).

\subsection{Comparisons with State-of-The-Arts}

\subsubsection{Comparison Methods}
We compare HOGS against a comprehensive set of state-of-the-art methods to evaluate its performance on both human-object and hand-object rendering.

For human-object rendering on HODome, comparison methods cover three main paradigms. 
\textbf{Mesh-based:} We include NeuralHumanFVV (NH-FVV)~\cite{suo2021neuralhumanfvv}, a representative method that learns to blend textured human reconstructions. 
\textbf{NeRF-based:} We compare against a suite of strong NeRF models, including the general novel-view synthesizer IBRNet~\cite{wang2021ibrnet}; 
NeuRay~\cite{liu2022neural}, which explicitly models visibility to handle occlusions; and NeuralHOIFVV (NHOI-FVV)~\cite{zhang2023neuraldome}, a state-of-the-art HOI-specific method using a layered NeRF representation.
\textbf{3DGS-based:} To create a challenging 3DGS competitor, we adapt \textbf{MANUS}~\cite{pokhariya2024manus}. 
As MANUS is originally designed for hands, we adapt it for full-body HOIs by replacing its hand component with our human deformation module while retaining its core object rendering pipeline.

For hand-object rendering on MANUS-Grasps, we compare HOGS against its original baseline, MANUS~\cite{pokhariya2024manus}, as well as two recent leading methods in hand-object rendering: HOLD~\cite{fan2024hold} and BIGS~\cite{on2025bigs}. 
This allows for a thorough evaluation of HOGS's extensibility and performance on fine-grained interactions.

\begin{table}[t]
\centering
\caption{\textbf{Quantitative results on the MANUS-Grasps dataset.} We evaluate HOGS against recent state-of-the-art methods MANUS, HOLD, and BIGS. We use \colorbox{red!20}{red} and \colorbox{yellow!40}{yellow} to denote the best and second-best results.}
\setlength{\tabcolsep}{12pt} 
\small
\begin{tabular}{lccc}
\toprule
Method & PSNR $\uparrow$ & SSIM $\uparrow$ & LPIPS $\downarrow$ \\
\midrule
HOLD~\cite{fan2024hold} & 25.913 & 0.9866 & 0.0722 \\
MANUS~\cite{pokhariya2024manus} & 26.328 & 0.9872 & \cellcolor{yellow!40}0.0688 \\
BIGS~\cite{on2025bigs} & \cellcolor{yellow!40}26.948 & \cellcolor{yellow!40}0.9885 & 0.0690 \\
\midrule
HOGS (Ours) & \cellcolor{red!20}27.425 & \cellcolor{red!20}0.9891 & \cellcolor{red!20}0.0651 \\
\bottomrule
\end{tabular}
\label{tab:manus_results}
\end{table}

\begin{figure}[t]
\centering
\includegraphics[width=0.45\textwidth]{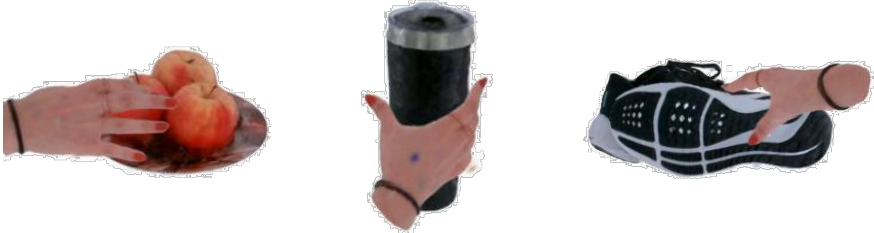} 
\caption{\textbf{Extensibility of HOGS.} Rendering results of diverse
hand-object grasping scenarios from the MANUS-Grasps dataset.}
\label{fig:extensiblity_qualitative}
\end{figure}

\subsubsection{Results on Human-object Rendering}
\noindent\textbf{Quantitative Analysis.}
Table~\ref{tab:quantitative} presents the quantitative comparison on the HODome dataset. 
Our proposed HOGS significantly outperforms all baselines across all evaluation metrics. 
Specifically, HOGS achieves a PSNR of 30.68 dB, surpassing the next-best 3DGS-based method by a large margin of 3.4 dB. 
This substantial improvement in rendering quality highlights the effectiveness of our composed Gaussian representation and physically plausible optimization.
In terms of efficiency, HOGS achieves a rendering speed of 162 FPS, which is not only the fastest among all methods but also orders of magnitude faster than NeRF-based approaches (1-2 FPS), demonstrating its suitability for real-time applications.

\noindent\textbf{Qualitative Analysis.}
Figure~\ref{fig:qualitative} presents qualitative comparisons of HOGS with existing methods. 
HOGS achieves superior rendering quality, generating novel views that closely match reference images. 
In contrast, the mesh-based NH-FVV struggles with occlusions, leading to noticeable artifacts and incomplete geometry. 
NeRF-based methods also exhibit various artifacts like blurry textures and geometric distortions. 
While the 3DGS-based MANUS captures overall human-object structure with reasonable rendering results, it lacks fine details in HOI regions.

\begin{table}[t]
\centering
\caption{\textbf{Ablation study of HOGS components.} We start with a baseline and incrementally add our proposed modules: LBS Modulation, Sparse-View (SV) Human Pose Refinement, Composed Gaussian Splatting, Physical loss, and SV Contact Prediction. We use \colorbox{red!20}{red} and \colorbox{yellow!40}{yellow} text to denote the best and second-best results of each metric respectively.}
    \centering
    \setlength{\tabcolsep}{5pt}
    \renewcommand{\arraystretch}{1.1}
    \small
    \begin{tabular}{lccc}
        \toprule
        Method & PSNR $\uparrow$ & SSIM $\uparrow$ & FPS $\uparrow$ \\
        \midrule
        Baseline & 26.96 & 0.928 & \cellcolor{red!20}179 \\
        $+$ LBS Modulation & 26.98 & 0.928 & \cellcolor{yellow!40}175 \\
        $+$ SV Human Pose Refinement & 27.40 & 0.931 & 174 \\
        $+$ Composed Gaussian Splatting & 29.12 & 0.944 & 171 \\
        $+$ Contact Loss & 29.95 & 0.951 & 165 \\
        $+$ Separation Loss & \cellcolor{red!20}30.85 & \cellcolor{red!20}0.956 & 157 \\
        $+$ SV Contact Prediction & \cellcolor{yellow!40}30.47 & \cellcolor{yellow!40}0.949 & 162 \\
        \bottomrule
    \end{tabular}
    \label{tab:ablation}
\end{table}

\subsubsection{Extensibility to Hand-Object Grasping}
\label{sec:exp_manus}
To demonstrate the flexibility and scalability of our framework, we extend HOGS to the hand-object grasping task and evaluate it on the MANUS-Grasps dataset. To adapt our method for this task, we substitute the SMPL-H full-body model in our human deformation stage (Sec.~\ref{sec:human_object_deformation}) with the widely-used MANO hand model~\cite{romero2017embodied}, while keeping all other components of our pipeline unchanged. This simple adaptation highlights the modularity of our design.

\noindent\textbf{Quantitative Analysis.} 
As shown in Table~\ref{tab:manus_results}, HOGS consistently achieves the best performance even when compared against state-of-the-art methods specifically designed for hand-object capture. Our method surpasses the recent strong baselines across all metrics, achieving a PSNR of 27.425 and an LPIPS score of 0.0651. This validates the effectiveness and strong generalization capability of our core pipeline for rendering complex, articulated interactions.

\noindent\textbf{Qualitative Analysis.} 
The visual results in Figure~\ref{fig:extensiblity_qualitative} further showcase HOGS's capability. Our method successfully renders a variety of challenging hand-object grasping scenes, capturing subtle details of finger-object contact and realistically representing diverse object shapes. 
These results confirm that our HOGS framework is not limited to full-body interactions but is broadly applicable to diverse articulated scenarios, demonstrating strong generalization capabilities across different scales of interaction.

\subsection{Component Analysis}
\label{exp:ablation}
In this section, we conduct a series of analytical experiments to validate the effectiveness of each key component within our HOGS framework.

\subsubsection{Ablation Study of Core Components}
We perform a comprehensive ablation study to demonstrate the cumulative contribution of our proposed modules. Starting from a basic setup, we incrementally add each component and report the performance on the HODome dataset (subject01 subset). The results are summarized in Table~\ref{tab:ablation}.

\noindent\textbf{Baseline Setup.}
Our baseline renders human and object via separate 3D Gaussian Splatting, then combines their point clouds for final rendering. 
In this baseline, human deformation solely relies on the LBS transformation (Sec.~\ref{sec:human_object_deformation}) with a randomly selected pose from the 6 input views as the target pose, while object deformation is performed consistently with the method detailed in Sec.~\ref{sec:human_object_deformation}.

\noindent\textbf{Incremental Module Analysis.}
We incrementally add our modules to the baseline, evaluating novel view synthesis. 
As shown in Table~\ref{tab:ablation}, rendering quality consistently improves with each added module, peaking with the Separation Loss. 
This progressive improvement demonstrates the effectiveness of our designed modules.

\begin{figure}[t]
    \centering
    \includegraphics[width=0.6\linewidth]{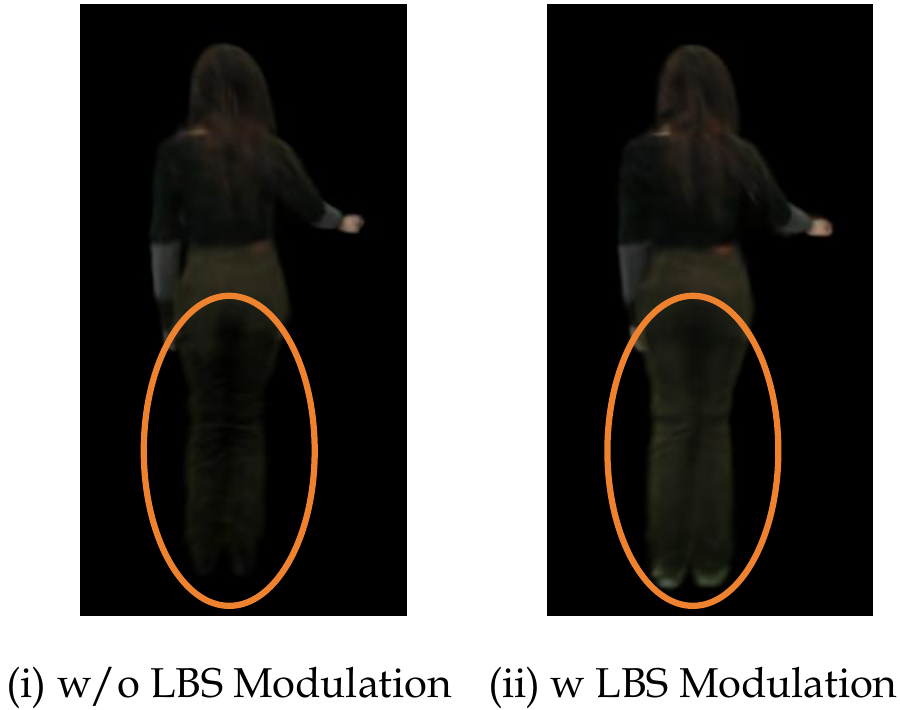} 
    \caption{\textbf{Visual ablation of LBS Modulation.} (i) Standard LBS results in over-smoothed, blurry boundaries. (ii) Our LBS Modulation recovers sharp details and corrects clothing geometry by breaking the linearity of SMPL.}
    \label{fig:lbs_ablation}
\end{figure}

\noindent\textbf{Impact of LBS Modulation.}
Regarding the LBS Modulation, although the quantitative gain in Table~\ref{tab:ablation} is subtle, its visual impact is critical. 
While standard LBS is computationally efficient, as a linear transformation, it is strictly limited to the underlying mesh topology, which often struggles to capture fine-grained details and subtle deformations essential for realistic rendering. 
Our LBS Modulation acts as a learnable residual field designed to break this linearity. As highlighted in Figure~\ref{fig:lbs_ablation}, standard LBS results in over-smoothed, blurry boundaries (e.g., around the legs). In contrast, our modulation successfully recovers sharp details and corrects clothing geometry.

\noindent\textbf{Impact of Sparse-View Contact Prediction Module.}
Incorporating sparse-view contact prediction yields a slight quality decrease but significantly boosts rendering efficiency (Table~\ref{tab:ablation}). 
This stems from focused physical loss computation. 
Without contact prediction, our \textit{Gaussian Contact and Separation Losses} (Eq.~\eqref{eq:contact_loss} and Eq.~\eqref{eq:separation_loss}) apply to all human-object Gaussian pairs---a thorough but computationally expensive process. 
However, HOI contact typically occurs in limited regions, involving partial Gaussians. 
Thus, sparse-view contact prediction is crucial for efficient HOI rendering, letting HOGS focus optimization on relevant contact regions to improve computational efficiency.

\begin{figure}[t]
    \centering
    \includegraphics[width=1.0\linewidth]{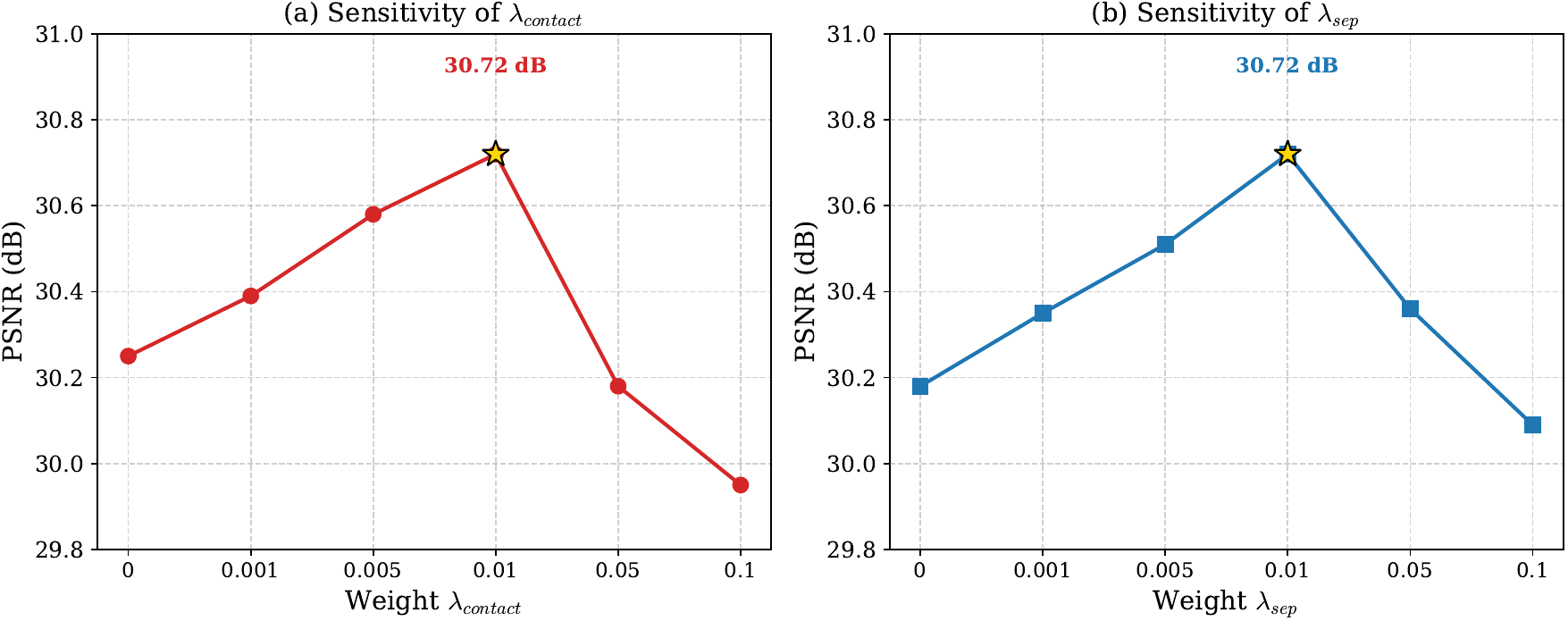} 
    \caption{\textbf{Sensitivity analysis of physical loss weights on the validation set.} Both (a) Contact Loss and (b) Separation Loss exhibit a clear inverted-U shaped trend, confirming $\lambda=0.01$ as the optimal balance point between geometric constraints and visual fidelity.}
    \label{fig:sensitivity_analysis}
\end{figure}

\noindent\textbf{Sensitivity of Physical Loss Weights.}
To validate our hyperparameter choice for physical constraints, we perform a sensitivity analysis on the validation set (Subject 08), as shown in Figure~\ref{fig:sensitivity_analysis}.
We employ a control variate method, varying one weight ($\lambda_{\text{contact}}$ or $\lambda_{\text{sep}}$) while fixing others.
Both losses exhibit a clear inverted-U shaped trend, peaking at 0.01.
Specifically, increasing $\lambda_{\text{contact}}$ up to 0.01 steadily improves PSNR by eliminating floating artifacts, but excessive weights ($>$0.05) degrade performance by unnaturally compressing geometry.
Similarly, $\lambda_{\text{sep}}=0.01$ provides the optimal balance for resolving inter-penetration without causing Gaussians to vanish (as observed with larger weights like 0.1).
This confirms that our chosen weights achieve the best trade-off between geometric plausibility and visual fidelity.

\begin{figure}[t] 
\centering 
\includegraphics[width=1.0\linewidth]{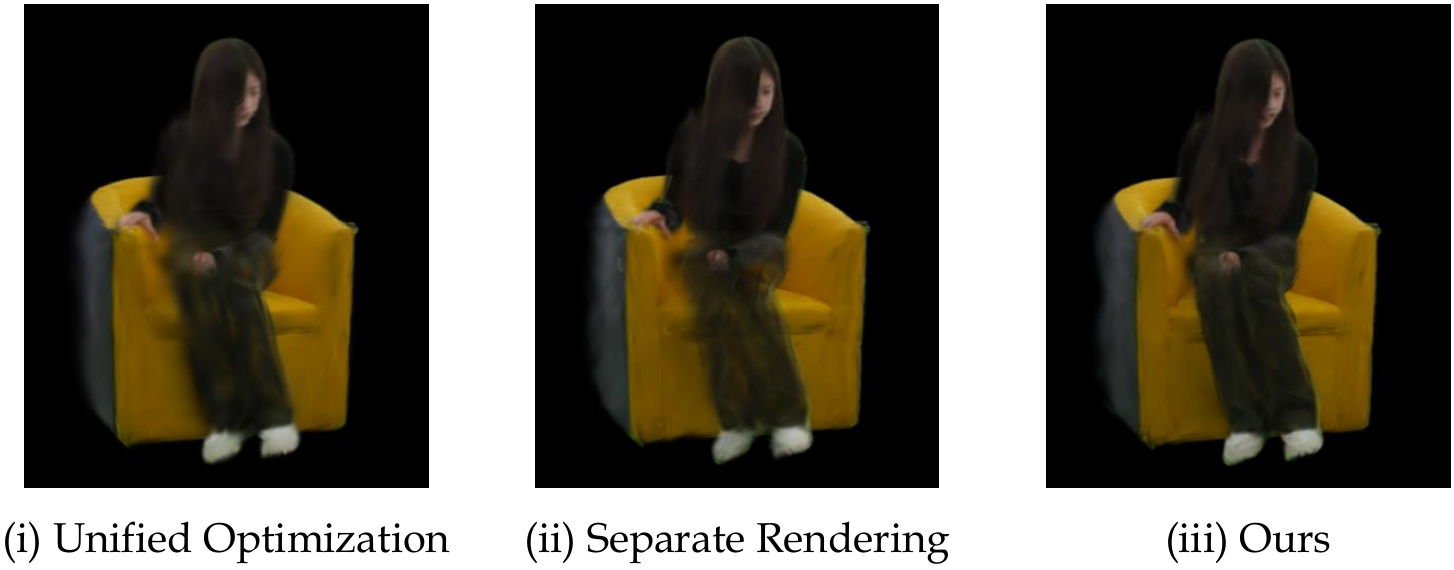} 
\caption{\textbf{Visual comparison of different rendering strategies.} (i) Unified Optimization suffers from severe artifacts and blurring on the rigid sofa. (ii) Separate Rendering leads to unnatural depth boundaries and edge artifacts. (iii) Ours preserves the sharp geometry of the rigid object while correctly handling the human-object depths.} 
\label{fig:composition_ablation} 
\end{figure}

\subsubsection{Analysis of Rendering Strategies}
\label{sec:ablation_strategy}

\begin{table}[t]
\centering
\small
\caption{\textbf{Comparisons of different human-object rendering strategies} on the HODome dataset.}
\label{tab:composition_ablation}
\setlength{\tabcolsep}{6pt} 
\begin{tabular}{lcc}
\toprule
Strategy & PSNR $\uparrow$ & SSIM $\uparrow$ \\
\midrule
Unified Optimization (Single Set) & 28.90 & 0.941 \\
Separate Rendering \& 2D Compositing & 28.47 & 0.935 \\
\midrule
\rowcolor{gray!20}\textbf{Ours (Composed Gaussian Splatting)} & \textbf{29.12} & \textbf{0.944} \\
\bottomrule
\end{tabular}
\end{table}

A core design choice in HOGS is \emph{Composed Gaussian Splatting}, which maintains separate Gaussian sets for human and object but renders them jointly. To verify the necessity of this strategy, we compare it against two alternatives: (1) \emph{\textbf{Unified Optimization}}, where human and object are treated as a single homogeneous Gaussian set; and (2) \emph{\textbf{Separate Rendering}}, where entities are rendered into separate images and composited using depth maps.

As shown in Table~\ref{tab:composition_ablation}, our strategy outperforms both baselines. Unified optimization suffers from \emph{dynamic conflict}—high gradients from the moving human cause artifacts on the rigid object (visualized in Figure~\ref{fig:composition_ablation}). Separate rendering fails to resolve complex occlusion boundaries in 3D space, leading to unnatural edges (-0.65 dB). Our approach effectively decouples the learning dynamics while ensuring correct visibility sorting, proving it is a vital component for high-fidelity HOI rendering.

\subsubsection{Effectiveness of Pre-trained Modules}
We provide qualitative results to further validate the two core pre-trained modules that enable robust performance from sparse views.

\begin{figure}[t]
\centering
\includegraphics[width=0.98\linewidth]{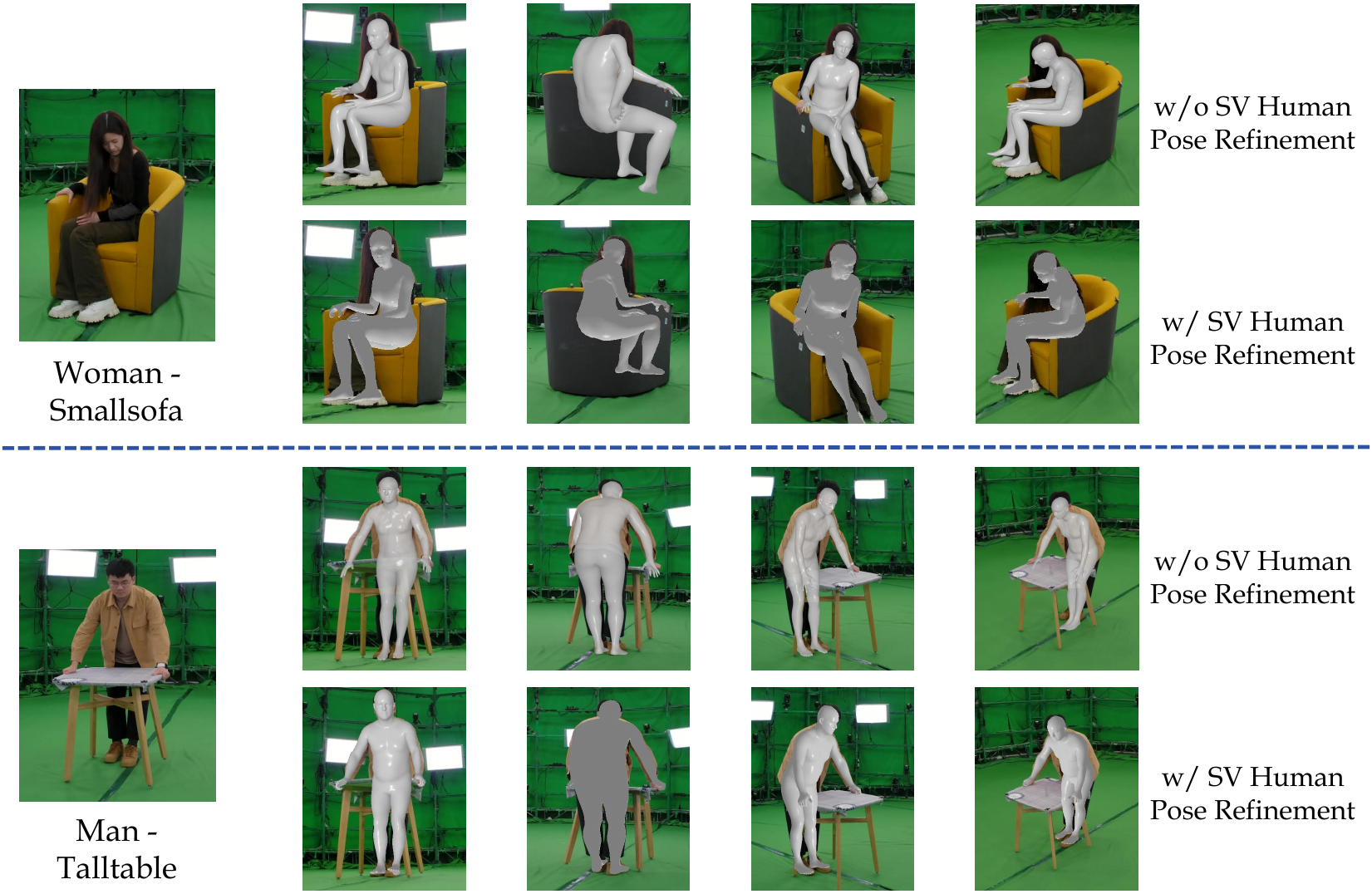} 
\caption{\textbf{Effectiveness of the SV Human Pose Refinement Module.} (a) Without refinement, using a pose from a single, potentially occluded view leads to incorrect body posture. (b) Our refined pose corrects these errors, resulting in a physically plausible and accurate human reconstruction.}
\label{fig:ablation_pose_refine}
\end{figure}

\begin{table}[t]
\centering
\small
\caption{\textbf{Comparison of Multi-view Human Pose Estimation Accuracy} (MPJPE/PA-MPJPE in mm) on the HODome dataset.}
\label{tab:refinement_comparison}
\setlength{\tabcolsep}{10pt}
\begin{tabular}{lcc}
\toprule
Method & MPJPE $\downarrow$ & PA-MPJPE $\downarrow$ \\
\midrule
Qiu et al.~\cite{Qiu_2019_ICCV} & 83.5 & 69.7 \\
AdaFuse~\cite{zhang2021adafuse}  & 71.0 & 61.6 \\
MvP~\cite{bragagnolo2024multi} & 71.3 & 61.4 \\
\midrule
Ours (Learnable $\alpha$) & 69.6 & 60.8 \\
\rowcolor{gray!20}\textbf{Ours (Fixed $\alpha=5$)} & \textbf{69.4} & \textbf{60.8} \\
\bottomrule
\end{tabular}
\end{table}

Figure~\ref{fig:ablation_pose_refine} illustrates the impact of our \textbf{\emph{human pose refinement}} module. 
As shown, relying on an off-the-shelf pose estimator~\cite{kanazawa2018end} from a single view can result in severe geometric artifacts due to occlusions (Fig.~\ref{fig:ablation_pose_refine}(a)). 
Our module effectively fuses information from all sparse views to produce a coherent and accurate 3D pose, leading to a much more realistic human rendering (Fig.~\ref{fig:ablation_pose_refine}(b)).
To further quantitatively validate our design, we compared our module against state-of-the-art multi-view pose estimation methods~\cite{Qiu_2019_ICCV, zhang2021adafuse, bragagnolo2024multi} on the HODome dataset. As reported in Table~\ref{tab:refinement_comparison}, our method achieves the lowest error with an MPJPE of 69.4 mm, outperforming the best baseline by a clear margin. Unlike standard fusion methods that treat views equally, our dynamic view weighting effectively filters out occluded noise, which is critical for robust HOI initialization. Moreover, we evaluate a variant where the sensitivity factor $\alpha$ is set as a learnable parameter. The learnable $\alpha$ yields a higher MPJPE than our fixed setting. This suggests that a fixed $\alpha$ effectively enforces the physical prior that occluded views are less reliable, preventing potential overfitting to noise in sparse-view inputs.

\begin{figure}[t]
\centering
\includegraphics[width=0.98\linewidth]{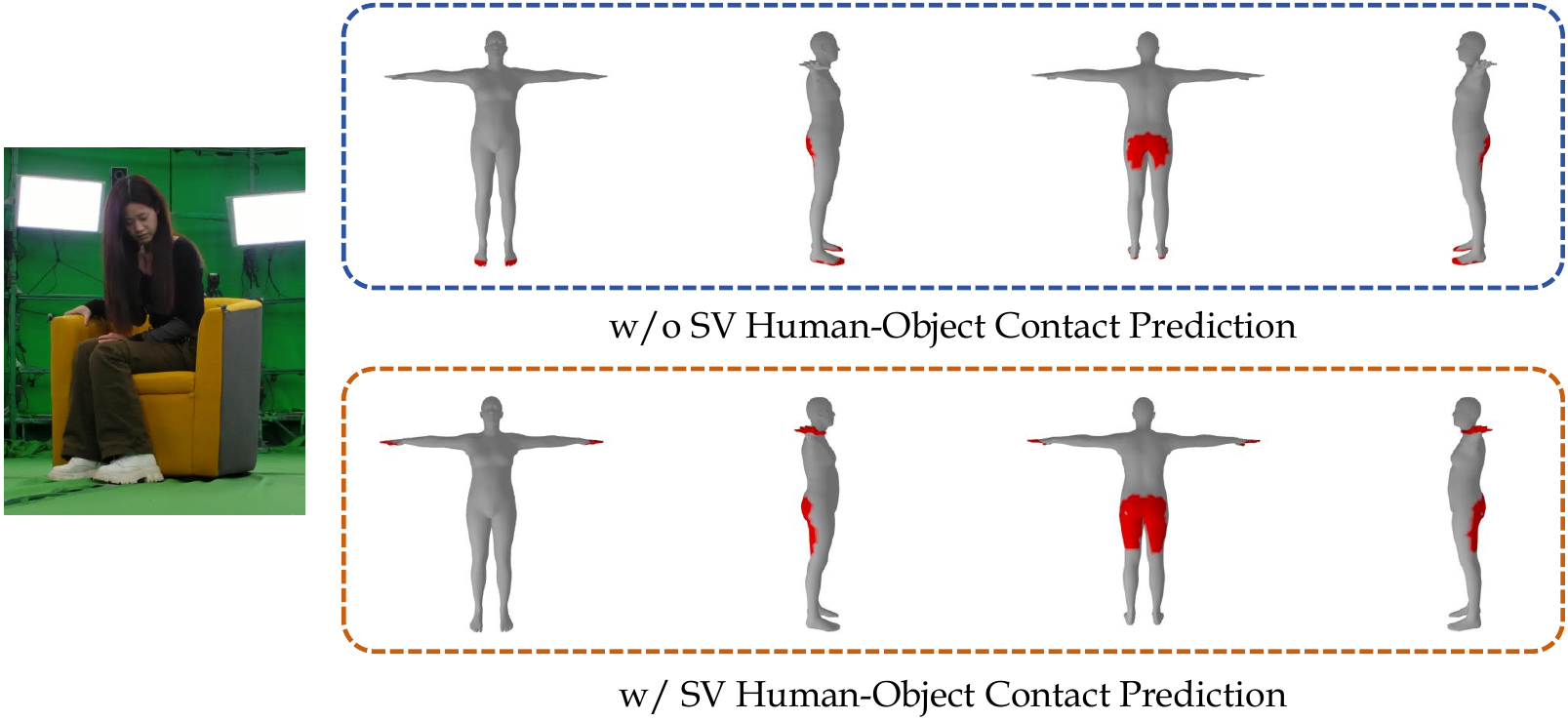} 
\caption{\textbf{Effectiveness of the SV Human-Object Contact Prediction Module.} Our module accurately identifies contact regions (highlighted in red) from sparse views, even for challenging interactions like sitting (left) and fine-grained hand contact (right), enabling focused and efficient physics optimization.}
\label{fig:ablation_cont_pred}
\end{figure}

Figure~\ref{fig:ablation_cont_pred} showcases the effectiveness of our \textbf{\emph{human-object contact prediction}} module. It can precisely locate the contact regions between the human and the object, from large contact surfaces (e.g., sitting on a sofa) to fine-grained interactions (e.g., hand touching an object). This accurate prediction serves as a critical prerequisite for ensuring the efficiency of our physically plausible optimization.

\subsubsection{Analysis of Physically Plausible Optimization}
To quantify the effect of our physical loss, we introduce the Mean Squared Penetration Depth (MSPD) metric. 
This metric averages the squared penetration depths of human Gaussian means (from predicted contact regions $\mathcal{C}$, see Sec.~\ref{sec:sparse_view_contact_prediction}) into the object's SDF. 
Specifically, MSPD is calculated as $\frac{1}{|\mathcal{C}|} \sum_{i \in \mathcal{C}} (\max(0, -\delta(\mu_i^h)))^2$, where $\delta(\mu_i^h)$ is the signed distance of the $i$-th human Gaussian mean $\mu_i^h$ within the object's SDF. 
A lower MSPD value indicates less penetration and thus better physical plausibility.
As detailed in Table~\ref{tab:effect_physical_losses}, applying our physical losses leads to a significant reduction in MSPD, decreasing the value from 3.829~cm$^2$ to 1.651~cm$^2$, which confirms their effectiveness in mitigating interpenetration.

\begin{figure}[t]
    \centering
    \includegraphics[width=0.7\linewidth]{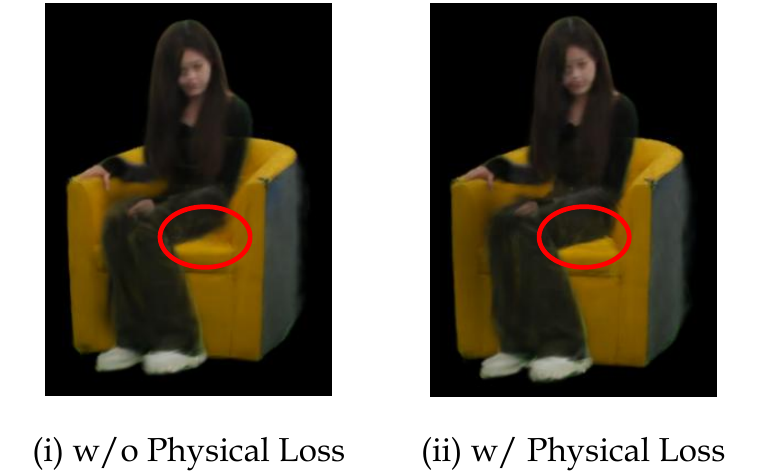} 
    \caption{\textbf{Qualitative effect of our physical losses}. (a) Without physical losses, the rendered human floats above the chair, showing a clear lack of physical contact. (b) With our losses, HOGS produces a plausible interaction where the person is sitting correctly on the chair.}
    \label{fig:ablation_visual}
\end{figure}

Complementing these quantitative findings, Figure~\ref{fig:ablation_visual} qualitatively demonstrates the impact of physical losses. 
As shown in Figure~\ref{fig:ablation_visual}(i), rendering without physical losses causes the human to float above the sofa, lacking plausible physical contact. 
Conversely, Figure~\ref{fig:ablation_visual}(ii) yields significant visual improvements. 
The rendered human now exhibits plausible contact with the chair, showcasing a more realistic and physically consistent interaction. 
This visual evidence, combined with the MSPD results, strongly validates the effectiveness of our proposed physical losses in enforcing plausible HOIs.

\begin{table}[t]
    \centering
    \setlength{\tabcolsep}{15pt} 
    \small
    \caption{\textbf{Quantitative effect of our physical losses}, measured by the Mean Squared Penetration Depth (MSPD).}
    \label{tab:effect_physical_losses}
    { 
    \begin{tabular}{cc}
        \toprule
        Method & MSPD (cm$^2$) $\downarrow$\\
        \midrule
        HOGS w/o Physical Losses & 3.829 \\
        HOGS w/ Physical Losses  & \textbf{1.651} \\
        \bottomrule
    \end{tabular}
    }
\end{table}

\subsection{Further Analysis}
\label{sec:further_analysis}

\begin{table}[t]
\centering
\small
\caption{\textbf{Quantitative robustness evaluation under extreme view sparsity.} Note that the results for the standard setting (6 views) are reported in Table~\ref{tab:quantitative}.}
\label{tab:view_sparsity}
\begin{tabular}{lcccc}
\toprule
\multirow{2}{*}{Method} & \multicolumn{2}{c}{2 Views} & \multicolumn{2}{c}{4 Views} \\
\cmidrule(lr){2-3} \cmidrule(lr){4-5} 
 & PSNR $\uparrow$ & SSIM $\uparrow$ & PSNR $\uparrow$ & SSIM $\uparrow$ \\
\midrule
NeuRay~\cite{liu2022neural} & 15.60 & 0.725 & 21.05 & 0.854 \\
MANUS~\cite{pokhariya2024manus} & 19.08 & 0.845 & 25.10 & 0.912 \\
\midrule
\rowcolor{gray!20}\textbf{HOGS (Ours)} & \textbf{23.36} & \textbf{0.887} & \textbf{29.15} & \textbf{0.942} \\
\bottomrule
\end{tabular}
\end{table}

\begin{figure}[t]
    \centering
    \includegraphics[width=0.75\linewidth]{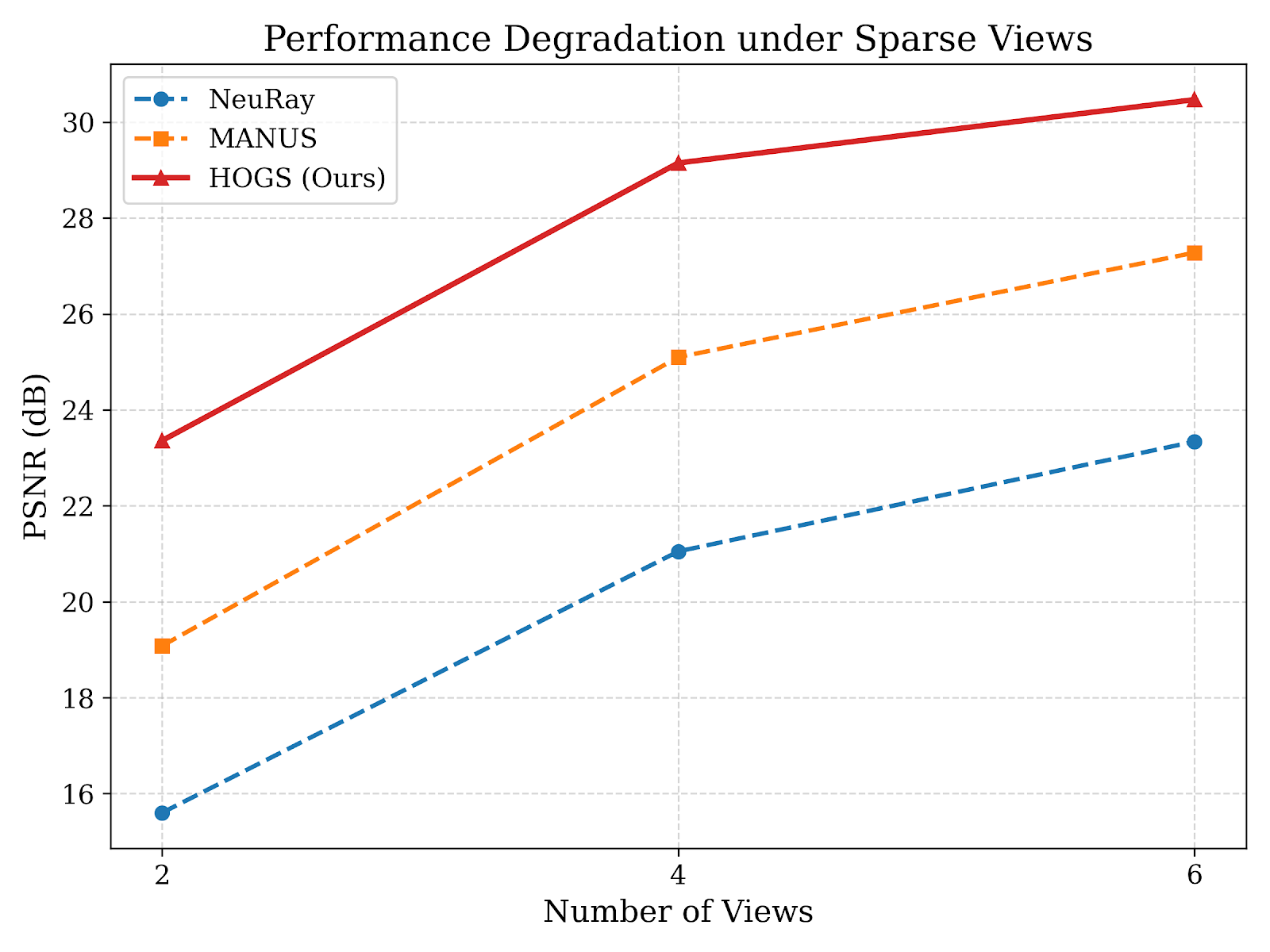} 
    \caption{\textbf{Performance degradation curve} under decreasing camera views.}
    \label{fig:view_ablation}
\end{figure}

\subsubsection{Robustness to Camera View Sparsity}
Existing sparse-view methods often degrade severely when the number of input views drops below a certain threshold. To evaluate the robustness of HOGS under extreme sparsity, we conduct a stress test by reducing the number of input views from 6 to 4 and 2. We compare our method against the leading NeRF-based method (NeuRay~\cite{liu2022neural}) and 3DGS-based method (MANUS~\cite{pokhariya2024manus}).

Table~\ref{tab:view_sparsity} presents the results under these extreme settings, while the standard 6-view performance serves as the baseline (refer to Table~\ref{tab:quantitative}). The degradation trends are visualized in Figure~\ref{fig:view_ablation}.
As observed, while all methods experience performance drops as views decrease, HOGS exhibits the most graceful degradation.
Specifically, at the extreme setting of 2 views, HOGS maintains a PSNR of 23.36 dB, significantly outperforming the 3DGS-based baseline MANUS (19.08 dB) by 4.28 dB.
Remarkably, by cross-referencing Table~\ref{tab:view_sparsity} with Table~\ref{tab:quantitative}, our performance at 2 views (23.36 dB) matches that of the NeRF-based NeuRay at 6 views (23.34 dB).
This resilience stems from our physically plausible optimization: when visual cues are critically lacking, the introduced geometric constraints (contact and separation) effectively prevent the geometric collapse that plagues baseline methods.

\subsubsection{Flexibility of Object Source}
The object deformation stage (Sec.~\ref{sec:human_object_deformation}) aims to acquire an object mesh, which serves as the basis for initializing object Gaussians. To align with NeuralDome~\cite{zhang2023neuraldome}, we directly employ its provided pre-scanned object templates. However, the requirement for pre-scanned templates is impractical for broader applications.

To assess HOGS's viability under more general conditions, we replace the template-based approach by reconstructing object meshes from sparse-view images using Pixel2Mesh++~\cite{wen2022pixel2mesh++}, thus unifying the input modality for both human and object deformations to sparse RGB images. 
Using these coarse reconstructed meshes naturally introduces geometric inaccuracies, which can affect the precision of initial Gaussian placement.
As shown in Table~\ref{tab:abl_obj_source}, this leads to a performance drop for all methods.

However, a comparative analysis highlights the superior robustness of our framework. When forced to use coarse reconstructed geometry, the state-of-the-art method MANUS~\cite{pokhariya2024manus} suffers a severe degradation of 2.43 dB. In contrast, HOGS demonstrates significantly better resilience with a smaller drop of 1.76 dB.
This suggests that our physically plausible optimization can better compensate for geometric initialization errors.
Crucially, even with these "noisy" reconstructed inputs, HOGS (28.71 dB) still outperforms the "perfect" version of the SOTA baseline (MANUS at 27.28 dB), confirming its practical viability in template-free scenarios.

\begin{table}[t]
    \centering
    \small
    \caption{\textbf{Robustness evaluation using different object mesh sources.} We compare HOGS against the SOTA method MANUS~\cite{pokhariya2024manus} using both high-fidelity templates and coarse meshes reconstructed from sparse views.}
    \label{tab:abl_obj_source}
    \setlength{\tabcolsep}{4pt} 
    \begin{tabular}{lccc}
        \toprule
        Method & Object Mesh Source & PSNR $\uparrow$ & Degradation $\downarrow$ \\
        \midrule
        MANUS [46] & Pre-scanned Template & 27.28 & - \\
        MANUS [46] & Reconstructed~\cite{wen2022pixel2mesh++} & 24.85 & -2.43 dB \\
        \midrule
        \textbf{HOGS (Ours)} & Pre-scanned Template & \textbf{30.47} & - \\
        \rowcolor{gray!20}\textbf{HOGS (Ours)} & Reconstructed~\cite{wen2022pixel2mesh++} & \textbf{28.71} & \textbf{-1.76 dB} \\
        \bottomrule
    \end{tabular}
\end{table}

\subsubsection{Robustness to Segmentation Mask Quality}
In our main experiments, we utilize reference masks provided by the HODome dataset, generated via Background Matting~\cite{sengupta2020background}. While accurate, this approach requires a pre-captured clean background, limiting in-the-wild applicability. To evaluate robustness and practical deployment potential, we adopt the Segment Anything Model (SAM)~\cite{kirillov2023segment} to extract masks using only coarse bounding-box prompts from the current frame.

\begin{figure}[t]
    \centering
    \includegraphics[width=0.8\linewidth]{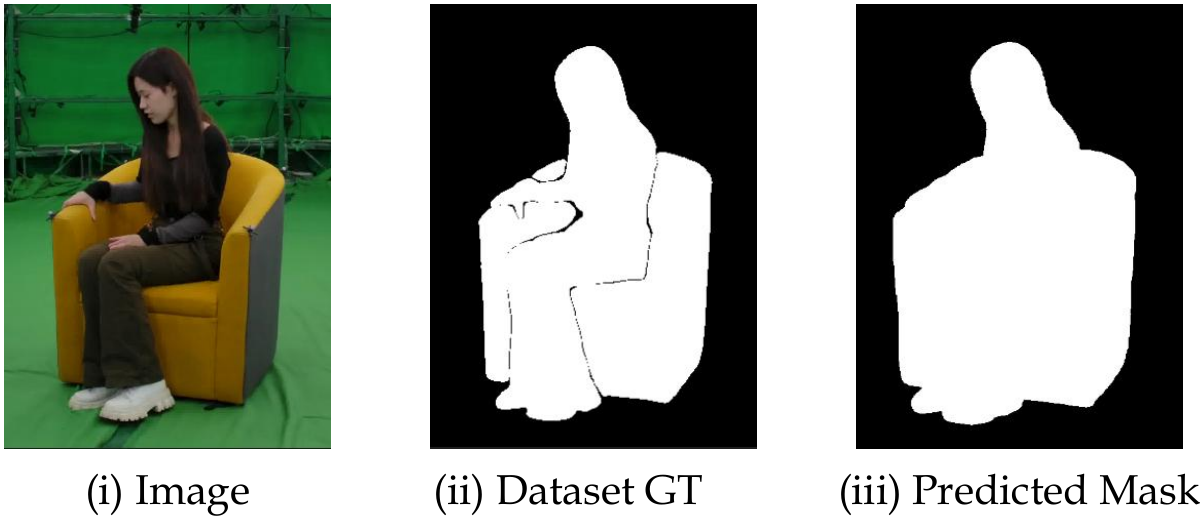}
    \caption{\textbf{Visual comparison of masks.} (i) Original Image. (ii) Dataset Reference exhibits gaps at the contact area. (iii) SAM predicts a continuous mask.}
    \label{fig:mask_comparison}
\end{figure}

We conduct an experiment on the Subject 01 subset by replacing dataset-provided masks entirely with SAM-predicted masks. As shown in Table~\ref{tab:mask_robustness}, using SAM-predicted masks yields comparable and slightly superior results to the original dataset reference.
The reason for this improvement is visualized in Figure~\ref{fig:mask_comparison}: the SAM-predicted mask is more faithful to physical reality, specifically depicting the human and object in full contact (Figure~\ref{fig:mask_comparison}(iii)), whereas background matting often leaves artifactual gaps (Figure~\ref{fig:mask_comparison}(ii)).
Crucially, the performance stability between these two distinct mask sources highlights the effectiveness of our physically plausible optimization. Our geometric constraints (Sec.~\ref{sec:physics_aware_optimization}) act as regularizers, allowing the model to recover plausible contact geometry even when input masks exhibit inconsistencies.

\begin{table}[t]
    \centering
    \small
    \caption{\textbf{Robustness evaluation on segmentation mask quality}. Comparison between dataset reference (Background Matting) and SAM-predicted masks.}
    \label{tab:mask_robustness}
    \begin{tabular}{llcc}
        \toprule
        Mask Source & Input Requirement & PSNR $\uparrow$ & SSIM $\uparrow$ \\
        \midrule
        HODome (Ref.) & Image + Background & 30.47 & 0.949 \\
        Predicted (SAM) & Image Only & 30.51 & 0.950 \\
        \bottomrule
    \end{tabular}
\end{table}

\begin{table}[t]
    \centering
    \small
    \caption{Efficiency comparison against adapted mesh-based contact strategies and global search baselines. Our prediction-guided SDF scheme achieves $\sim$17$\times$ speedup.}
    \label{tab:optimization_efficiency}
    \begin{tabular}{llcc}
        \toprule
        Category & Method & Time / Iter. (s) $\downarrow$ & Speedup $\uparrow$ \\
        \midrule
        \multirow{3}{*}{\shortstack[l]{Mesh-Based\\(Adapted)}} 
        & BEHAVE~\cite{bhatnagar2022behave} & 6.259 & 1.0$\times$ \\
        & CHORE~\cite{xie2022chore} & 6.527 & 1.0$\times$ \\
        & GraviCap~\cite{dabral2021gravity} & 6.325 & 1.0$\times$ \\
        \midrule
        \multirow{2}{*}{\shortstack[l]{3DGS-Based}}
        & Global Search & 5.722 & 1.1$\times$ \\
        & \textbf{Ours} & \textbf{0.375} & \textbf{17.4$\times$} \\
        \bottomrule
    \end{tabular}
\end{table}

\subsubsection{Efficiency and Training Time}
Our framework is designed for both high-quality rendering and efficiency. As shown in Table~\ref{tab:training_times}, the \emph{Sparse-View Human Pose Refinement} and \emph{Sparse-View Human-Object Contact Prediction} modules are pre-trained offline as a one-time cost ($\sim$15 and $\sim$24 minutes, respectively). Subsequently, for each HOI scene, the main HOGS pipeline performs per-scene optimization.

To validate the computational advantage of our scheme, we compare HOGS against two baselines: (1) \textbf{Adapted Mesh-based Methods}, where we implement contact losses from representative works (BEHAVE~\cite{bhatnagar2022behave}, CHORE~\cite{xie2022chore}, GraviCap~\cite{dabral2021gravity}) using Nearest Neighbor search adapted for Gaussians; and (2) \textbf{Global Search}, an ablation of our method that computes SDF queries for all Gaussians without contact prediction.

As shown in Table~\ref{tab:optimization_efficiency}, the adapted mesh-based methods are computationally expensive ($\sim$6.3s per iteration) due to the complexity of neighbor searching in unstructured Gaussian fields. Even using efficient SDF queries (Global Search) only reduces this to 5.7s. 
In contrast, by combining SDF queries with our predicted contact mask, HOGS narrows the optimization scope significantly, reducing the physics calculation to just \textbf{0.375s per iteration}.
This achieves a speedup of over \textbf{16$\times$} compared to standard approaches, confirming that our design is critical for enabling real-time, physically plausible optimization.

This efficiency is critical for the overall pipeline. Thanks to this accelerated physics computation, HOGS's main pipeline requires only 7.5 minutes of total optimization time per scene. This is notably more efficient than the baseline MANUS~\cite{pokhariya2024manus}, which requires 9 minutes. This highlights HOGS's advantage in rapid adaptation for novel HOI sequences.

\begin{table}[t]
    \centering
    \caption{\textbf{Training and optimization times.} Pre-training is a one-time cost. Per-scene optimization is for adapting to a new sequence.}
    \label{tab:training_times}
    \renewcommand{\arraystretch}{1.1}
    \setlength{\tabcolsep}{15pt} 
    \small
    \begin{tabular}{@{}lc@{}} 
        \toprule
        Method / Component & Time (min) \\
        \midrule 
        HOGS: Human Pose Refinement (Pre-train) & $\sim$15 \\ 
        HOGS: Contact Prediction (Pre-train)    & $\sim$24 \\ 
        \midrule
        HOGS: Main Pipeline (Per-Scene Optim.)  & 7.5 \\ 
        MANUS~\cite{pokhariya2024manus} (Per-Scene Optim.) & 9 \\
        \bottomrule
    \end{tabular}
\end{table}

%% file: sec/5_conclusion.tex
\section{Conclusion}

In this paper, we present HOGS, a novel approach for rendering realistic and physically plausible HOIs from sparse views. 
Its key innovation is the integration of composed Gaussian Splatting with physically plausible rendering optimization, supported by sparse-view human pose refinement and human-object contact prediction. 
This combination enables HOGS to effectively capture intricate human-object interplay, generating high-quality novel views that adhere to physical constraints. 
Extensive experiments demonstrate HOGS achieves superior performance in rendering quality, efficiency, and physical plausibility over existing methods, while also showing broader applicability to articulated object interactions.

\noindent\textbf{Limitations and Future Work.}
While HOGS marks a significant step forward for HOI rendering, the current framework mainly focuses on interactions between a single human and a single object. 
This is a common setup in foundational HOI research, but extending it to more complex scenes is a crucial next step. We consider that our component-based design provides a clear path for such extensions. 
One could extend HOGS to multi-person or multi-object scenarios by creating multiple instances of our human/object deformation module and introducing inter-entity physics (e.g., human-human or object-object collision) into the optimization. 
Furthermore, to handle non-rigid objects, the current ICP-based object deformation module could be replaced with more advanced models for articulated or deformable objects, while keeping the core physically plausible rendering pipeline intact.

%% file: main.bib
@String(ICCV= {Int. Conf. Comput. Vis.})

@String(TOG= {ACM Trans. Graph.})

@String(ICIP = {IEEE Int. Conf. Image Process.})

@String(AAAI = {AAAI})

@String(ICCV  = {ICCV})

@String(TOG   = {ACM TOG})

@String(ICIP  = {ICIP})

@article{gupta2009observing,
  title={Observing human-object interactions: Using spatial and functional compatibility for recognition},
  author={Gupta, Abhinav and Kembhavi, Aniruddha and Davis, Larry S},
  journal={IEEE transactions on pattern analysis and machine intelligence},
  volume={31},
  number={10},
  pages={1775--1789},
  year={2009},
  publisher={IEEE}
}

@inproceedings{hassan2021populating,
  title={Populating 3D scenes by learning human-scene interaction},
  author={Hassan, Mohamed and Ghosh, Partha and Tesch, Joachim and Tzionas, Dimitrios and Black, Michael J},
  booktitle={Proceedings of the IEEE/CVF Conference on Computer Vision and Pattern Recognition},
  pages={14708--14718},
  year={2021}
}

@inproceedings{huang2022capturing,
  title={Capturing and inferring dense full-body human-scene contact},
  author={Huang, Chun-Hao P and Yi, Hongwei and H{\"o}schle, Markus and Safroshkin, Matvey and Alexiadis, Tsvetelina and Polikovsky, Senya and Scharstein, Daniel and Black, Michael J},
  booktitle={Proceedings of the IEEE/CVF Conference on Computer Vision and Pattern Recognition},
  pages={13274--13285},
  year={2022}
}

@inproceedings{liu2025revisit,
  title={Revisit human-scene interaction via space occupancy},
  author={Liu, Xinpeng and Hou, Haowen and Yang, Yanchao and Li, Yong-Lu and Lu, Cewu},
  booktitle={European Conference on Computer Vision},
  pages={1--19},
  year={2025},
  organization={Springer}
}

@inproceedings{li2024task,
  title={Task-oriented human-object interactions generation with implicit neural representations},
  author={Li, Quanzhou and Wang, Jingbo and Loy, Chen Change and Dai, Bo},
  booktitle={Proceedings of the IEEE/CVF Winter Conference on Applications of Computer Vision},
  pages={3035--3044},
  year={2024}
}

@article{fernandez2020associated,
  title={Associated reality: A cognitive human--machine layer for autonomous driving},
  author={Fernandez, Felipe and Sanchez, Angel and Velez, Jose F and Moreno, Belen},
  journal={Robotics and Autonomous Systems},
  volume={133},
  pages={103624},
  year={2020},
  publisher={Elsevier}
}

@article{kong2022human,
  title={Human action recognition and prediction: A survey},
  author={Kong, Yu and Fu, Yun},
  journal={International Journal of Computer Vision},
  volume={130},
  number={5},
  pages={1366--1401},
  year={2022},
  publisher={Springer}
}

@inproceedings{zhao2024m,
  title={I'M HOI: Inertia-aware Monocular Capture of 3D Human-Object Interactions},
  author={Zhao, Chengfeng and Zhang, Juze and Du, Jiashen and Shan, Ziwei and Wang, Junye and Yu, Jingyi and Wang, Jingya and Xu, Lan},
  booktitle={Proceedings of the IEEE/CVF Conference on Computer Vision and Pattern Recognition},
  pages={729--741},
  year={2024}
}

@inproceedings{pang2024sparse,
  title={Sparse multi-view hand-object reconstruction for unseen environments},
  author={Pang, Yik Lung and Oh, Changjae and Cavallaro, Andrea},
  booktitle={Proceedings of the IEEE/CVF Conference on Computer Vision and Pattern Recognition},
  pages={803--810},
  year={2024}
}

@inproceedings{jiang2022neuralhofusion,
  title={Neuralhofusion: Neural volumetric rendering under human-object interactions},
  author={Jiang, Yuheng and Jiang, Suyi and Sun, Guoxing and Su, Zhuo and Guo, Kaiwen and Wu, Minye and Yu, Jingyi and Xu, Lan},
  booktitle={Proceedings of the IEEE/CVF Conference on Computer Vision and Pattern Recognition},
  pages={6155--6165},
  year={2022}
}

@inproceedings{sun2021neural,
  title={Neural free-viewpoint performance rendering under complex human-object interactions},
  author={Sun, Guoxing and Chen, Xin and Chen, Yizhang and Pang, Anqi and Lin, Pei and Jiang, Yuheng and Xu, Lan and Yu, Jingyi and Wang, Jingya},
  booktitle={Proceedings of the 29th ACM International Conference on Multimedia},
  pages={4651--4660},
  year={2021}
}

@inproceedings{jiang2023instant,
  title={Instant-NVR: Instant Neural Volumetric Rendering for Human-object Interactions from Monocular RGBD Stream},
  author={Jiang, Yuheng and Yao, Kaixin and Su, Zhuo and Shen, Zhehao and Luo, Haimin and Xu, Lan},
  booktitle={Proceedings of the IEEE/CVF Conference on Computer Vision and Pattern Recognition},
  pages={595--605},
  year={2023}
}

@inproceedings{zhang2023neuraldome,
  title={NeuralDome: A neural modeling pipeline on multi-view human-object interactions},
  author={Zhang, Juze and Luo, Haimin and Yang, Hongdi and Xu, Xinru and Wu, Qianyang and Shi, Ye and Yu, Jingyi and Xu, Lan and Wang, Jingya},
  booktitle={Proceedings of the IEEE/CVF Conference on Computer Vision and Pattern Recognition},
  pages={8834--8845},
  year={2023}
}

@inproceedings{bhatnagar2022behave,
  title={Behave: Dataset and method for tracking human object interactions},
  author={Bhatnagar, Bharat Lal and Xie, Xianghui and Petrov, Ilya A and Sminchisescu, Cristian and Theobalt, Christian and Pons-Moll, Gerard},
  booktitle={Proceedings of the IEEE/CVF Conference on Computer Vision and Pattern Recognition},
  pages={15935--15946},
  year={2022}
}

@inproceedings{hassan2019resolving,
  title={Resolving 3D human pose ambiguities with 3D scene constraints},
  author={Hassan, Mohamed and Choutas, Vasileios and Tzionas, Dimitrios and Black, Michael J},
  booktitle={Proceedings of the IEEE/CVF international conference on computer vision},
  pages={2282--2292},
  year={2019}
}

@inproceedings{taheri2020grab,
  title={GRAB: A dataset of whole-body human grasping of objects},
  author={Taheri, Omid and Ghorbani, Nima and Black, Michael J and Tzionas, Dimitrios},
  booktitle={Computer Vision--ECCV 2020: 16th European Conference, Glasgow, UK, August 23--28, 2020, Proceedings, Part IV 16},
  pages={581--600},
  year={2020},
  organization={Springer}
}

@article{collet2015high,
  title={High-quality streamable free-viewpoint video},
  author={Collet, Alvaro and Chuang, Ming and Sweeney, Pat and Gillett, Don and Evseev, Dennis and Calabrese, David and Hoppe, Hugues and Kirk, Adam and Sullivan, Steve},
  journal={ACM Transactions on Graphics (ToG)},
  volume={34},
  number={4},
  pages={1--13},
  year={2015},
  publisher={ACM New York, NY, USA}
}

@article{dou2017motion2fusion,
  title={Motion2fusion: Real-time volumetric performance capture},
  author={Dou, Mingsong and Davidson, Philip and Fanello, Sean Ryan and Khamis, Sameh and Kowdle, Adarsh and Rhemann, Christoph and Tankovich, Vladimir and Izadi, Shahram},
  journal={ACM Transactions on Graphics (ToG)},
  volume={36},
  number={6},
  pages={1--16},
  year={2017},
  publisher={ACM New York, NY, USA}
}

@inproceedings{schonberger2016structure,
  title={Structure-from-motion revisited},
  author={Schonberger, Johannes L and Frahm, Jan-Michael},
  booktitle={Proceedings of the IEEE conference on computer vision and pattern recognition},
  pages={4104--4113},
  year={2016}
}

@article{mildenhall2021nerf,
  title={Nerf: Representing scenes as neural radiance fields for view synthesis},
  author={Mildenhall, Ben and Srinivasan, Pratul P and Tancik, Matthew and Barron, Jonathan T and Ramamoorthi, Ravi and Ng, Ren},
  journal={Communications of the ACM},
  volume={65},
  number={1},
  pages={99--106},
  year={2021},
  publisher={ACM New York, NY, USA}
}

@article{kerbl20233d,
  title={3d gaussian splatting for real-time radiance field rendering.},
  author={Kerbl, Bernhard and Kopanas, Georgios and Leimk{\"u}hler, Thomas and Drettakis, George},
  journal={ACM Trans. Graph.},
  volume={42},
  number={4},
  pages={139--1},
  year={2023}
}

@article{fei20243d,
  title={3d gaussian splatting as new era: A survey},
  author={Fei, Ben and Xu, Jingyi and Zhang, Rui and Zhou, Qingyuan and Yang, Weidong and He, Ying},
  journal={IEEE Transactions on Visualization and Computer Graphics},
  year={2024},
  publisher={IEEE}
}

@inproceedings{shaw2025swings,
  title={Swings: sliding windows for dynamic 3D gaussian splatting},
  author={Shaw, Richard and Nazarczuk, Michal and Song, Jifei and Moreau, Arthur and Catley-Chandar, Sibi and Dhamo, Helisa and P{\'e}rez-Pellitero, Eduardo},
  booktitle={European Conference on Computer Vision},
  pages={37--54},
  year={2025},
  organization={Springer}
}

@inproceedings{liu2024humangaussian,
  title={Humangaussian: Text-driven 3d human generation with gaussian splatting},
  author={Liu, Xian and Zhan, Xiaohang and Tang, Jiaxiang and Shan, Ying and Zeng, Gang and Lin, Dahua and Liu, Xihui and Liu, Ziwei},
  booktitle={Proceedings of the IEEE/CVF Conference on Computer Vision and Pattern Recognition},
  pages={6646--6657},
  year={2024}
}

@inproceedings{moreau2024human,
  title={Human gaussian splatting: Real-time rendering of animatable avatars},
  author={Moreau, Arthur and Song, Jifei and Dhamo, Helisa and Shaw, Richard and Zhou, Yiren and P{\'e}rez-Pellitero, Eduardo},
  booktitle={Proceedings of the IEEE/CVF Conference on Computer Vision and Pattern Recognition},
  pages={788--798},
  year={2024}
}

@inproceedings{hu2024gauhuman,
  title={Gauhuman: Articulated gaussian splatting from monocular human videos},
  author={Hu, Shoukang and Hu, Tao and Liu, Ziwei},
  booktitle={Proceedings of the IEEE/CVF Conference on Computer Vision and Pattern Recognition},
  pages={20418--20431},
  year={2024}
}

@inproceedings{qian20243dgs,
  title={3dgs-avatar: Animatable avatars via deformable 3d gaussian splatting},
  author={Qian, Zhiyin and Wang, Shaofei and Mihajlovic, Marko and Geiger, Andreas and Tang, Siyu},
  booktitle={Proceedings of the IEEE/CVF Conference on Computer Vision and Pattern Recognition},
  pages={5020--5030},
  year={2024}
}

@inproceedings{kocabas2024hugs,
  title={Hugs: Human gaussian splats},
  author={Kocabas, Muhammed and Chang, Jen-Hao Rick and Gabriel, James and Tuzel, Oncel and Ranjan, Anurag},
  booktitle={Proceedings of the IEEE/CVF conference on computer vision and pattern recognition},
  pages={505--515},
  year={2024}
}

@article{romero2022embodied,
  title={Embodied hands: Modeling and capturing hands and bodies together},
  author={Romero, Javier and Tzionas, Dimitrios and Black, Michael J},
  journal={arXiv preprint arXiv:2201.02610},
  year={2022}
}

@inproceedings{tripathi2023deco,
  title={DECO: Dense estimation of 3D human-scene contact in the wild},
  author={Tripathi, Shashank and Chatterjee, Agniv and Passy, Jean-Claude and Yi, Hongwei and Tzionas, Dimitrios and Black, Michael J},
  booktitle={Proceedings of the IEEE/CVF International Conference on Computer Vision},
  pages={8001--8013},
  year={2023}
}

@inproceedings{smolic2004free,
  title={Free viewpoint video extraction, representation, coding, and rendering},
  author={Smolic, Aljoscha and Mueller, Karsten and Merkle, Philipp and Rein, Tobias and Kautzner, Matthias and Eisert, Peter and Wiegand, Thomas},
  booktitle={2004 International Conference on Image Processing, 2004. ICIP'04.},
  volume={5},
  pages={3287--3290},
  year={2004},
  organization={IEEE}
}

@inproceedings{bansal20204d,
  title={4d visualization of dynamic events from unconstrained multi-view videos},
  author={Bansal, Aayush and Vo, Minh and Sheikh, Yaser and Ramanan, Deva and Narasimhan, Srinivasa},
  booktitle={Proceedings of the IEEE/CVF Conference on Computer Vision and Pattern Recognition},
  pages={5366--5375},
  year={2020}
}

@article{shum2000review,
  title={Review of image-based rendering techniques},
  author={Shum, Harry and Kang, Sing Bing},
  journal={Visual Communications and Image Processing 2000},
  volume={4067},
  pages={2--13},
  year={2000},
  publisher={SPIE}
}

@inproceedings{zhi2021place,
  title={In-place scene labelling and understanding with implicit scene representation},
  author={Zhi, Shuaifeng and Laidlow, Tristan and Leutenegger, Stefan and Davison, Andrew J},
  booktitle={Proceedings of the IEEE/CVF International Conference on Computer Vision},
  pages={15838--15847},
  year={2021}
}

@inproceedings{wu2022object,
  title={Object-compositional neural implicit surfaces},
  author={Wu, Qianyi and Liu, Xian and Chen, Yuedong and Li, Kejie and Zheng, Chuanxia and Cai, Jianfei and Zheng, Jianmin},
  booktitle={European Conference on Computer Vision},
  pages={197--213},
  year={2022},
  organization={Springer}
}

@article{ding2024point,
  title={point diffusion implicit function for large-scale scene neural representation},
  author={Ding, Yuhan and Yin, Fukun and Fan, Jiayuan and Li, Hui and Chen, Xin and Liu, Wen and Lu, Chongshan and Yu, Gang and Chen, Tao},
  journal={Advances in Neural Information Processing Systems},
  volume={36},
  year={2024}
}

@inproceedings{cao2023hexplane,
  title={Hexplane: A fast representation for dynamic scenes},
  author={Cao, Ang and Johnson, Justin},
  booktitle={Proceedings of the IEEE/CVF Conference on Computer Vision and Pattern Recognition},
  pages={130--141},
  year={2023}
}

@inproceedings{shetty2024holoported,
  title={Holoported Characters: Real-time Free-viewpoint Rendering of Humans from Sparse RGB Cameras},
  author={Shetty, Ashwath and Habermann, Marc and Sun, Guoxing and Luvizon, Diogo and Golyanik, Vladislav and Theobalt, Christian},
  booktitle={Proceedings of the IEEE/CVF Conference on Computer Vision and Pattern Recognition},
  pages={1206--1215},
  year={2024}
}

@inproceedings{weng2022humannerf,
  title={Humannerf: Free-viewpoint rendering of moving people from monocular video},
  author={Weng, Chung-Yi and Curless, Brian and Srinivasan, Pratul P and Barron, Jonathan T and Kemelmacher-Shlizerman, Ira},
  booktitle={Proceedings of the IEEE/CVF conference on computer vision and pattern Recognition},
  pages={16210--16220},
  year={2022}
}

@article{yang2022neural,
  title={Neural rendering in a room: amodal 3d understanding and free-viewpoint rendering for the closed scene composed of pre-captured objects},
  author={Yang, Bangbang and Zhang, Yinda and Li, Yijin and Cui, Zhaopeng and Fanello, Sean and Bao, Hujun and Zhang, Guofeng},
  journal={ACM Transactions on Graphics (TOG)},
  volume={41},
  number={4},
  pages={1--10},
  year={2022},
  publisher={ACM New York, NY, USA}
}

@inproceedings{xian2021space,
  title={Space-time neural irradiance fields for free-viewpoint video},
  author={Xian, Wenqi and Huang, Jia-Bin and Kopf, Johannes and Kim, Changil},
  booktitle={Proceedings of the IEEE/CVF conference on computer vision and pattern recognition},
  pages={9421--9431},
  year={2021}
}

@inproceedings{jayasundara2023flexnerf,
  title={FlexNeRF: Photorealistic free-viewpoint rendering of moving humans from sparse views},
  author={Jayasundara, Vinoj and Agrawal, Amit and Heron, Nicolas and Shrivastava, Abhinav and Davis, Larry S},
  booktitle={Proceedings of the IEEE/CVF Conference on Computer Vision and Pattern Recognition},
  pages={21118--21127},
  year={2023}
}

@inproceedings{zhang2025cor,
  title={CoR-GS: sparse-view 3D Gaussian splatting via co-regularization},
  author={Zhang, Jiawei and Li, Jiahe and Yu, Xiaohan and Huang, Lei and Gu, Lin and Zheng, Jin and Bai, Xiao},
  booktitle={European Conference on Computer Vision},
  pages={335--352},
  year={2025},
  organization={Springer}
}

@inproceedings{xu2024relightable,
  title={Relightable and animatable neural avatar from sparse-view video},
  author={Xu, Zhen and Peng, Sida and Geng, Chen and Mou, Linzhan and Yan, Zihan and Sun, Jiaming and Bao, Hujun and Zhou, Xiaowei},
  booktitle={Proceedings of the IEEE/CVF Conference on Computer Vision and Pattern Recognition},
  pages={990--1000},
  year={2024}
}

@inproceedings{kwon2025generalizable,
  title={Generalizable human gaussians for sparse view synthesis},
  author={Kwon, Youngjoong and Fang, Baole and Lu, Yixing and Dong, Haoye and Zhang, Cheng and Carrasco, Francisco Vicente and Mosella-Montoro, Albert and Xu, Jianjin and Takagi, Shingo and Kim, Daeil and others},
  booktitle={European Conference on Computer Vision},
  pages={451--468},
  year={2025},
  organization={Springer}
}

@inproceedings{petersen2022gendr,
  title={Gendr: A generalized differentiable renderer},
  author={Petersen, Felix and Goldluecke, Bastian and Borgelt, Christian and Deussen, Oliver},
  booktitle={Proceedings of the IEEE/CVF Conference on Computer Vision and Pattern Recognition},
  pages={4002--4011},
  year={2022}
}

@inproceedings{bangaru2022differentiable,
  title={Differentiable rendering of neural sdfs through reparameterization},
  author={Bangaru, Sai Praveen and Gharbi, Michael and Luan, Fujun and Li, Tzu-Mao and Sunkavalli, Kalyan and Hasan, Milos and Bi, Sai and Xu, Zexiang and Bernstein, Gilbert and Durand, Fredo},
  booktitle={SIGGRAPH Asia 2022 Conference Papers},
  pages={1--9},
  year={2022}
}

@article{worchel2023differentiable,
  title={Differentiable Rendering of Parametric Geometry},
  author={Worchel, Markus and Alexa, Marc},
  journal={ACM Transactions on Graphics (TOG)},
  volume={42},
  number={6},
  pages={1--18},
  year={2023},
  publisher={ACM New York, NY, USA}
}

@article{muller2022instant,
  title={Instant neural graphics primitives with a multiresolution hash encoding},
  author={M{\"u}ller, Thomas and Evans, Alex and Schied, Christoph and Keller, Alexander},
  journal={ACM transactions on graphics (TOG)},
  volume={41},
  number={4},
  pages={1--15},
  year={2022},
  publisher={ACM New York, NY, USA}
}

@inproceedings{zhu2025fsgs,
  title={Fsgs: Real-time few-shot view synthesis using gaussian splatting},
  author={Zhu, Zehao and Fan, Zhiwen and Jiang, Yifan and Wang, Zhangyang},
  booktitle={European conference on computer vision},
  pages={145--163},
  year={2025},
  organization={Springer}
}

@inproceedings{liu2025citygaussian,
  title={Citygaussian: Real-time high-quality large-scale scene rendering with gaussians},
  author={Liu, Yang and Luo, Chuanchen and Fan, Lue and Wang, Naiyan and Peng, Junran and Zhang, Zhaoxiang},
  booktitle={European Conference on Computer Vision},
  pages={265--282},
  year={2025},
  organization={Springer}
}

@inproceedings{zhang2024hoi,
  title={HOI-M\^{} 3: Capture Multiple Humans and Objects Interaction within Contextual Environment},
  author={Zhang, Juze and Zhang, Jingyan and Song, Zining and Shi, Zhanhe and Zhao, Chengfeng and Shi, Ye and Yu, Jingyi and Xu, Lan and Wang, Jingya},
  booktitle={Proceedings of the IEEE/CVF Conference on Computer Vision and Pattern Recognition},
  pages={516--526},
  year={2024}
}

@inproceedings{hu2024hand,
  title={Hand-Object Interaction Controller (HOIC): Deep Reinforcement Learning for Reconstructing Interactions with Physics},
  author={Hu, Haoyu and Yi, Xinyu and Cao, Zhe and Yong, Jun-Hai and Xu, Feng},
  booktitle={ACM SIGGRAPH 2024 Conference Papers},
  pages={1--10},
  year={2024}
}

@article{battaglia2016interaction,
  title={Interaction networks for learning about objects, relations and physics},
  author={Battaglia, Peter and Pascanu, Razvan and Lai, Matthew and Jimenez Rezende, Danilo and others},
  journal={Advances in neural information processing systems},
  volume={29},
  year={2016}
}

@inproceedings{jain2009interactive,
  title={Interactive synthesis of human-object interaction},
  author={Jain, Sumit and Liu, C Karen},
  booktitle={Proceedings of the 2009 ACM SIGGRAPH/Eurographics Symposium on Computer Animation},
  pages={47--53},
  year={2009}
}

@inproceedings{huang2020arch,
  title={Arch: Animatable reconstruction of clothed humans},
  author={Huang, Zeng and Xu, Yuanlu and Lassner, Christoph and Li, Hao and Tung, Tony},
  booktitle={Proceedings of the IEEE/CVF Conference on Computer Vision and Pattern Recognition},
  pages={3093--3102},
  year={2020}
}

@inproceedings{peng2021animatable,
  title={Animatable neural radiance fields for modeling dynamic human bodies},
  author={Peng, Sida and Dong, Junting and Wang, Qianqian and Zhang, Shangzhan and Shuai, Qing and Zhou, Xiaowei and Bao, Hujun},
  booktitle={Proceedings of the IEEE/CVF International Conference on Computer Vision},
  pages={14314--14323},
  year={2021}
}

@inproceedings{lin2022learning,
  title={Learning implicit templates for point-based clothed human modeling},
  author={Lin, Siyou and Zhang, Hongwen and Zheng, Zerong and Shao, Ruizhi and Liu, Yebin},
  booktitle={European Conference on Computer Vision},
  pages={210--228},
  year={2022},
  organization={Springer}
}

@inproceedings{liu2023hosnerf,
  title={Hosnerf: Dynamic human-object-scene neural radiance fields from a single video},
  author={Liu, Jia-Wei and Cao, Yan-Pei and Yang, Tianyuan and Xu, Zhongcong and Keppo, Jussi and Shan, Ying and Qie, Xiaohu and Shou, Mike Zheng},
  booktitle={Proceedings of the IEEE/CVF International Conference on Computer Vision},
  pages={18483--18494},
  year={2023}
}

@inproceedings{kanazawa2018end,
  title={End-to-end recovery of human shape and pose},
  author={Kanazawa, Angjoo and Black, Michael J and Jacobs, David W and Malik, Jitendra},
  booktitle={Proceedings of the IEEE conference on computer vision and pattern recognition},
  pages={7122--7131},
  year={2018}
}

@inproceedings{bogo2016keep,
  title={Keep it SMPL: Automatic estimation of 3D human pose and shape from a single image},
  author={Bogo, Federica and Kanazawa, Angjoo and Lassner, Christoph and Gehler, Peter and Romero, Javier and Black, Michael J},
  booktitle={Computer Vision--ECCV 2016: 14th European Conference, Amsterdam, The Netherlands, October 11-14, 2016, Proceedings, Part V 14},
  pages={561--578},
  year={2016},
  organization={Springer}
}

@inproceedings{besl1992method,
  title={Method for registration of 3-D shapes},
  author={Besl, Paul J and McKay, Neil D},
  booktitle={Sensor fusion IV: control paradigms and data structures},
  volume={1611},
  pages={586--606},
  year={1992},
  organization={Spie}
}

@article{wang2004image,
  title={Image quality assessment: from error visibility to structural similarity},
  author={Wang, Zhou and Bovik, Alan C and Sheikh, Hamid R and Simoncelli, Eero P},
  journal={IEEE transactions on image processing},
  volume={13},
  number={4},
  pages={600--612},
  year={2004},
  publisher={IEEE}
}

@inproceedings{zhang2018unreasonable,
  title={The unreasonable effectiveness of deep features as a perceptual metric},
  author={Zhang, Richard and Isola, Phillip and Efros, Alexei A and Shechtman, Eli and Wang, Oliver},
  booktitle={Proceedings of the IEEE conference on computer vision and pattern recognition},
  pages={586--595},
  year={2018}
}

@inproceedings{li2019parametric,
  title={Parametric model-based 3D human shape and pose estimation from multiple views},
  author={Li, Zhongguo and Heyden, Anders and Oskarsson, Magnus},
  booktitle={Image Analysis: 21st Scandinavian Conference, SCIA 2019, Norrk{\"o}ping, Sweden, June 11--13, 2019, Proceedings 21},
  pages={336--347},
  year={2019},
  organization={Springer}
}

@inproceedings{paliwal2025coherentgs,
  title={Coherentgs: Sparse novel view synthesis with coherent 3d gaussians},
  author={Paliwal, Avinash and Ye, Wei and Xiong, Jinhui and Kotovenko, Dmytro and Ranjan, Rakesh and Chandra, Vikas and Kalantari, Nima Khademi},
  booktitle={European Conference on Computer Vision},
  pages={19--37},
  year={2025},
  organization={Springer}
}

@inproceedings{chen2025mvsplat,
  title={Mvsplat: Efficient 3d gaussian splatting from sparse multi-view images},
  author={Chen, Yuedong and Xu, Haofei and Zheng, Chuanxia and Zhuang, Bohan and Pollefeys, Marc and Geiger, Andreas and Cham, Tat-Jen and Cai, Jianfei},
  booktitle={European Conference on Computer Vision},
  pages={370--386},
  year={2025},
  organization={Springer}
}

@inproceedings{mihajlovic2025splatfields,
  title={Splatfields: Neural gaussian splats for sparse 3d and 4d reconstruction},
  author={Mihajlovic, Marko and Prokudin, Sergey and Tang, Siyu and Maier, Robert and Bogo, Federica and Tung, Tony and Boyer, Edmond},
  booktitle={European Conference on Computer Vision},
  pages={313--332},
  year={2025},
  organization={Springer}
}

@inproceedings{pokhariya2024manus,
  title={MANUS: Markerless Grasp Capture using Articulated 3D Gaussians},
  author={Pokhariya, Chandradeep and Shah, Ishaan Nikhil and Xing, Angela and Li, Zekun and Chen, Kefan and Sharma, Avinash and Sridhar, Srinath},
  booktitle={Proceedings of the IEEE/CVF Conference on Computer Vision and Pattern Recognition},
  pages={2197--2208},
  year={2024}
}

@article{romero2017embodied,
  title={Embodied hands: modeling and capturing hands and bodies together},
  author={Romero, Javier and Tzionas, Dimitrios and Black, Michael J},
  journal={ACM Transactions on Graphics (TOG)},
  volume={36},
  number={6},
  pages={1--17},
  year={2017},
  publisher={ACM New York, NY, USA}
}

@inproceedings{suo2021neuralhumanfvv,
  title={Neuralhumanfvv: Real-time neural volumetric human performance rendering using rgb cameras},
  author={Suo, Xin and Jiang, Yuheng and Lin, Pei and Zhang, Yingliang and Wu, Minye and Guo, Kaiwen and Xu, Lan},
  booktitle={Proceedings of the IEEE/CVF conference on computer vision and pattern recognition},
  pages={6226--6237},
  year={2021}
}

@inproceedings{wang2021ibrnet,
  title={Ibrnet: Learning multi-view image-based rendering},
  author={Wang, Qianqian and Wang, Zhicheng and Genova, Kyle and Srinivasan, Pratul P and Zhou, Howard and Barron, Jonathan T and Martin-Brualla, Ricardo and Snavely, Noah and Funkhouser, Thomas},
  booktitle={Proceedings of the IEEE/CVF conference on computer vision and pattern recognition},
  pages={4690--4699},
  year={2021}
}

@inproceedings{liu2022neural,
  title={Neural rays for occlusion-aware image-based rendering},
  author={Liu, Yuan and Peng, Sida and Liu, Lingjie and Wang, Qianqian and Wang, Peng and Theobalt, Christian and Zhou, Xiaowei and Wang, Wenping},
  booktitle={Proceedings of the IEEE/CVF Conference on Computer Vision and Pattern Recognition},
  pages={7824--7833},
  year={2022}
}

@inproceedings{zhou2025nerfect,
  title={The nerfect match: Exploring nerf features for visual localization},
  author={Zhou, Qunjie and Maximov, Maxim and Litany, Or and Leal-Taix{\'e}, Laura},
  booktitle={European Conference on Computer Vision},
  pages={108--127},
  year={2025},
  organization={Springer}
}

@inproceedings{li2024gp,
  title={GP-NeRF: Generalized Perception NeRF for Context-Aware 3D Scene Understanding},
  author={Li, Hao and Zhang, Dingwen and Dai, Yalun and Liu, Nian and Cheng, Lechao and Li, Jingfeng and Wang, Jingdong and Han, Junwei},
  booktitle={Proceedings of the IEEE/CVF Conference on Computer Vision and Pattern Recognition},
  pages={21708--21718},
  year={2024}
}

@inproceedings{tonderski2024neurad,
  title={Neurad: Neural rendering for autonomous driving},
  author={Tonderski, Adam and Lindstr{\"o}m, Carl and Hess, Georg and Ljungbergh, William and Svensson, Lennart and Petersson, Christoffer},
  booktitle={Proceedings of the IEEE/CVF Conference on Computer Vision and Pattern Recognition},
  pages={14895--14904},
  year={2024}
}

@inproceedings{boos2016flashback,
  title={Flashback: Immersive virtual reality on mobile devices via rendering memoization},
  author={Boos, Kevin and Chu, David and Cuervo, Eduardo},
  booktitle={Proceedings of the 14th Annual International Conference on Mobile Systems, Applications, and Services},
  pages={291--304},
  year={2016}
}

@article{qiao2019web,
  title={Web AR: A promising future for mobile augmented reality—State of the art, challenges, and insights},
  author={Qiao, Xiuquan and Ren, Pei and Dustdar, Schahram and Liu, Ling and Ma, Huadong and Chen, Junliang},
  journal={Proceedings of the IEEE},
  volume={107},
  number={4},
  pages={651--666},
  year={2019},
  publisher={IEEE}
}

@inproceedings{yang2021cpf,
  title={Cpf: Learning a contact potential field to model the hand-object interaction},
  author={Yang, Lixin and Zhan, Xinyu and Li, Kailin and Xu, Wenqiang and Li, Jiefeng and Lu, Cewu},
  booktitle={Proceedings of the IEEE/CVF International Conference on Computer Vision},
  pages={11097--11106},
  year={2021}
}

@inproceedings{xie2023visibility,
  title={Visibility aware human-object interaction tracking from single rgb camera},
  author={Xie, Xianghui and Bhatnagar, Bharat Lal and Pons-Moll, Gerard},
  booktitle={Proceedings of the IEEE/CVF Conference on Computer Vision and Pattern Recognition},
  pages={4757--4768},
  year={2023}
}

@article{su2022robustfusion,
  title={Robustfusion: Robust volumetric performance reconstruction under human-object interactions from monocular rgbd stream},
  author={Su, Zhuo and Xu, Lan and Zhong, Dawei and Li, Zhong and Deng, Fan and Quan, Shuxue and Fang, Lu},
  journal={IEEE Transactions on Pattern Analysis and Machine Intelligence},
  volume={45},
  number={5},
  pages={6196--6213},
  year={2022},
  publisher={IEEE}
}

@article{wen2022pixel2mesh++,
  title={Pixel2mesh++: 3d mesh generation and refinement from multi-view images},
  author={Wen, Chao and Zhang, Yinda and Cao, Chenjie and Li, Zhuwen and Xue, Xiangyang and Fu, Yanwei},
  journal={IEEE Transactions on Pattern Analysis and Machine Intelligence},
  volume={45},
  number={2},
  pages={2166--2180},
  year={2022},
  publisher={IEEE}
}

@inproceedings{xie2022chore,
  title={Chore: Contact, human and object reconstruction from a single rgb image},
  author={Xie, Xianghui and Bhatnagar, Bharat Lal and Pons-Moll, Gerard},
  booktitle={European Conference on Computer Vision},
  pages={125--145},
  year={2022},
  organization={Springer}
}

@inproceedings{jiang2023full,
  title={Full-body articulated human-object interaction},
  author={Jiang, Nan and Liu, Tengyu and Cao, Zhexuan and Cui, Jieming and Zhang, Zhiyuan and Chen, Yixin and Wang, He and Zhu, Yixin and Huang, Siyuan},
  booktitle={Proceedings of the IEEE/CVF International Conference on Computer Vision},
  pages={9365--9376},
  year={2023}
}

@inproceedings{feng2025flashgs,
  title={Flashgs: Efficient 3d gaussian splatting for large-scale and high-resolution rendering},
  author={Feng, Guofeng and Chen, Siyan and Fu, Rong and Liao, Zimu and Wang, Yi and Liu, Tao and Hu, Boni and Xu, Linning and Pei, Zhilin and Li, Hengjie and others},
  booktitle={Proceedings of the Computer Vision and Pattern Recognition Conference},
  pages={26652--26662},
  year={2025}
}

@inproceedings{fan2024hold,
  title={Hold: Category-agnostic 3d reconstruction of interacting hands and objects from video},
  author={Fan, Zicong and Parelli, Maria and Kadoglou, Maria Eleni and Chen, Xu and Kocabas, Muhammed and Black, Michael J and Hilliges, Otmar},
  booktitle={Proceedings of the IEEE/CVF Conference on Computer Vision and Pattern Recognition},
  pages={494--504},
  year={2024}
}

@inproceedings{on2025bigs,
  title={BIGS: Bimanual Category-agnostic Interaction Reconstruction from Monocular Videos via 3D Gaussian Splatting},
  author={On, Jeongwan and Gwak, Kyeonghwan and Kang, Gunyoung and Cha, Junuk and Hwang, Soohyun and Hwang, Hyein and Baek, Seungryul},
  booktitle={Proceedings of the Computer Vision and Pattern Recognition Conference},
  pages={17437--17447},
  year={2025}
}

@inproceedings{liu2024citygaussian,
  title={Citygaussian: Real-time high-quality large-scale scene rendering with gaussians},
  author={Liu, Yang and Luo, Chuanchen and Fan, Lue and Wang, Naiyan and Peng, Junran and Zhang, Zhaoxiang},
  booktitle={European Conference on Computer Vision},
  pages={265--282},
  year={2024},
  organization={Springer}
}

@inproceedings{kong2025efficient,
  title={Efficient gaussian splatting for monocular dynamic scene rendering via sparse time-variant attribute modeling},
  author={Kong, Hanyang and Yang, Xingyi and Wang, Xinchao},
  booktitle={Proceedings of the AAAI Conference on Artificial Intelligence},
  volume={39},
  number={4},
  pages={4374--4382},
  year={2025}
}

@inproceedings{li2023multi,
  title={Multi-view inverse rendering for large-scale real-world indoor scenes},
  author={Li, Zhen and Wang, Lingli and Cheng, Mofang and Pan, Cihui and Yang, Jiaqi},
  booktitle={Proceedings of the IEEE/CVF Conference on Computer Vision and Pattern Recognition},
  pages={12499--12509},
  year={2023}
}

@inproceedings{moon2024expressive,
  title={Expressive whole-body 3d gaussian avatar},
  author={Moon, Gyeongsik and Shiratori, Takaaki and Saito, Shunsuke},
  booktitle={European Conference on Computer Vision},
  pages={19--35},
  year={2024},
  organization={Springer}
}

@inproceedings{qiu2025anigs,
  title={Anigs: Animatable gaussian avatar from a single image with inconsistent gaussian reconstruction},
  author={Qiu, Lingteng and Zhu, Shenhao and Zuo, Qi and Gu, Xiaodong and Dong, Yuan and Zhang, Junfei and Xu, Chao and Li, Zhe and Yuan, Weihao and Bo, Liefeng and others},
  booktitle={Proceedings of the Computer Vision and Pattern Recognition Conference},
  pages={21148--21158},
  year={2025}
}

@inproceedings{pumarola2021d,
  title={D-nerf: Neural radiance fields for dynamic scenes},
  author={Pumarola, Albert and Corona, Enric and Pons-Moll, Gerard and Moreno-Noguer, Francesc},
  booktitle={Proceedings of the IEEE/CVF conference on computer vision and pattern recognition},
  pages={10318--10327},
  year={2021}
}

@article{su2021nerf,
  title={A-nerf: Articulated neural radiance fields for learning human shape, appearance, and pose},
  author={Su, Shih-Yang and Yu, Frank and Zollh{\"o}fer, Michael and Rhodin, Helge},
  journal={Advances in neural information processing systems},
  volume={34},
  pages={12278--12291},
  year={2021}
}

@ARTICLE{11024125,
  author={Bao, Zhenyu and Liao, Guibiao and Zhou, Kaichen and Liu, Kanglin and Li, Qing and Qiu, Guoping},
  journal={IEEE Transactions on Image Processing}, 
  title={LoopSparseGS: Loop-Based Sparse-View Friendly Gaussian Splatting}, 
  year={2025},
  volume={34},
  number={},
  pages={3889-3902},
  doi={10.1109/TIP.2025.3574929}}

@inproceedings{sun2025real,
  title={Real-time Free-view Human Rendering from Sparse-view RGB Videos using Double Unprojected Textures},
  author={Sun, Guoxing and Dabral, Rishabh and Zhu, Heming and Fua, Pascal and Theobalt, Christian and Habermann, Marc},
  booktitle={Proceedings of the Computer Vision and Pattern Recognition Conference},
  pages={562--573},
  year={2025}
}

@article{zhou2023hdhuman,
  title={Hdhuman: High-quality human novel-view rendering from sparse views},
  author={Zhou, Tiansong and Huang, Jing and Yu, Tao and Shao, Ruizhi and Li, Kun},
  journal={IEEE Transactions on Visualization and Computer Graphics},
  volume={30},
  number={8},
  pages={5328--5338},
  year={2023},
  publisher={IEEE}
}

@InProceedings{Qiu_2019_ICCV,
author = {Qiu, Haibo and Wang, Chunyu and Wang, Jingdong and Wang, Naiyan and Zeng, Wenjun},
title = {Cross View Fusion for 3D Human Pose Estimation},
booktitle = {Proceedings of the IEEE/CVF International Conference on Computer Vision (ICCV)},
month = {October},
year = {2019}
}

@inproceedings{he2020epipolar,
  title={Epipolar transformer for multi-view human pose estimation},
  author={He, Yihui and Yan, Rui and Fragkiadaki, Katerina and Yu, Shoou-I},
  booktitle={Proceedings of the IEEE/CVF conference on computer vision and pattern recognition workshops},
  pages={1036--1037},
  year={2020}
}

@inproceedings{iskakov2019learnable,
  title={Learnable triangulation of human pose},
  author={Iskakov, Karim and Burkov, Egor and Lempitsky, Victor and Malkov, Yury},
  booktitle={Proceedings of the IEEE/CVF international conference on computer vision},
  pages={7718--7727},
  year={2019}
}

@article{ma2021transfusion,
  title={Transfusion: Cross-view fusion with transformer for 3d human pose estimation},
  author={Ma, Haoyu and Chen, Liangjian and Kong, Deying and Wang, Zhe and Liu, Xingwei and Tang, Hao and Yan, Xiangyi and Xie, Yusheng and Lin, Shih-Yao and Xie, Xiaohui},
  journal={arXiv preprint arXiv:2110.09554},
  year={2021}
}

@article{wan2023view,
  title={View consistency aware holistic triangulation for 3d human pose estimation},
  author={Wan, Xiaoyue and Chen, Zhuo and Zhao, Xu},
  journal={Computer Vision and Image Understanding},
  volume={236},
  pages={103830},
  year={2023},
  publisher={Elsevier}
}

@article{zhang2021adafuse,
  title={Adafuse: Adaptive multiview fusion for accurate human pose estimation in the wild},
  author={Zhang, Zhe and Wang, Chunyu and Qiu, Weichao and Qin, Wenhu and Zeng, Wenjun},
  journal={International Journal of Computer Vision},
  volume={129},
  number={3},
  pages={703--718},
  year={2021},
  publisher={Springer}
}

@inproceedings{bragagnolo2024multi,
  title={Multi-view Pose Fusion for Occlusion-Aware 3D Human Pose Estimation},
  author={Bragagnolo, Laura and Terreran, Matteo and Allegro, Davide and Ghidoni, Stefano},
  booktitle={European Conference on Computer Vision},
  pages={117--133},
  year={2024},
  organization={Springer}
}

@inproceedings{dabral2021gravity,
  title={Gravity-aware monocular 3d human-object reconstruction},
  author={Dabral, Rishabh and Shimada, Soshi and Jain, Arjun and Theobalt, Christian and Golyanik, Vladislav},
  booktitle={Proceedings of the IEEE/CVF International Conference on Computer Vision},
  pages={12365--12374},
  year={2021}
}

@inproceedings{sengupta2020background,
  title={Background matting: The world is your green screen},
  author={Sengupta, Soumyadip and Jayaram, Vivek and Curless, Brian and Seitz, Steven M and Kemelmacher-Shlizerman, Ira},
  booktitle={Proceedings of the IEEE/CVF Conference on Computer Vision and Pattern Recognition},
  pages={2291--2300},
  year={2020}
}

@inproceedings{kirillov2023segment,
  title={Segment anything},
  author={Kirillov, Alexander and Mintun, Eric and Ravi, Nikhila and Mao, Hanzi and Rolland, Chloe and Gustafson, Laura and Xiao, Tete and Whitehead, Spencer and Berg, Alexander C and Lo, Wan-Yen and others},
  booktitle={Proceedings of the IEEE/CVF international conference on computer vision},
  pages={4015--4026},
  year={2023}
}
